\documentclass{jfm}

\usepackage{array}
\usepackage{epsfig}
\usepackage{epsf}
\usepackage{amssymb}

\def\be{\mbox{\boldmath $e$}}
\def\bu{\mbox{\boldmath $u$}}
\def\bOmega{\mbox{\boldmath $\Omega$}}
\def\dd{{\rm d}}
\def\grad{\nabla}
\def\ptl{\partial}
\def\ou{{\bar u}}

\title[Turbulent shear flows]{A model for the nonlinear dynamics of
  turbulent shear flows}

\author[P. Garaud \& G. I. Ogilvie]{P\ls A\ls S\ls C\ls A\ls
  L\ls E\ns G\ls A\ls R\ls A\ls U\ls D \and G\ls O\ls R\ls D\ls O\ls
  N\ns I.\ns O\ls G\ls I\ls L\ls V\ls I\ls E}

\affiliation{Institute of Astronomy, University of Cambridge,
  Madingley Road, Cambridge CB3 0HA, UK\\[\affilskip]
  Department of Applied Mathematics and Theoretical Physics, Centre
  for Mathematical Sciences, University of Cambridge, Wilberforce
  Road, Cambridge CB3 0WA, UK}

\begin{document}

\maketitle

\begin{abstract}
  We investigate the nonlinear dynamics of turbulent shear flows, with
  and without rotation, in the context of a simple but physically
  motivated closure of the equation governing the evolution of the
  Reynolds stress tensor.  We show that the model naturally accounts
  for some familiar phenomena in parallel shear flows, such as the
  subcritical transition to turbulence at a finite Reynolds number and
  the occurrence of a universal velocity profile close to a wall at
  large Reynolds number. For rotating shear flows we find that,
  depending on the Rayleigh discriminant of the system, the model
  predicts either linear instability or nonlinear instability or
  complete stability as the Reynolds number is increased to large
  values. We investigate the properties of Couette--Taylor flows for
  varying inner and outer cylinder rotation rates and identify the
  region of linear instability (similar to Taylor's), as well as
  regions of finite-amplitude instability qualitatively compatible
  with recent experiments. We also discuss quantitative predictions of
  the model in comparison with a range of experimental torque
  measurements.  Finally, we consider the relevance of this work to
  the question of the hydrodynamic stability of astrophysical
  accretion discs.
\end{abstract}

\section{Introduction}

Turbulent motion in the flow of an incompressible fluid has
consistently eluded a satisfying mathematical description in over a
century of investigation.  The governing Navier--Stokes equations,
although involving only a relatively simple nonlinearity, conceal a
remarkable wealth of complex behaviour.  Even in the simplest problem
of the onset of turbulent motion in a parallel shear flow, the
classical approach based on a linear analysis of normal modes fails
spectacularly to account for the basic experimental results (e.g.
Drazin \& Reid 1981).  A better insight into the transition to
turbulence has been gained more recently through studies of the
transient amplification of disturbances in linear theory (Butler \&
Farrell 1992), which, when coupled with an appropriate nonlinear
feedback, allows perturbations to be sustained (e.g. Baggett \&
Trefethen 1997).  The direct computation of nonlinear disturbances in
the form of (possibly unstable) steady solutions or travelling waves
that act as precursors of the turbulent dynamics (Nagata 1990; Waleffe
1997) has also shed light on the process of transition.

By its nature, fully developed turbulence demands a statistical
description. Reynolds (1895) established the principles of a
statistical theory, showing that correlations between components of
the fluctuating velocity field provide a stress that influences the
bulk motion.  The analogy between this turbulent transport of momentum
and the viscous transport associated with thermal molecular motion in
kinetic theory suggested the concept of eddy viscosity, introduced by
Boussinesq (1877).  Later, Prandtl's (1925) theory of the mixing
length gave a predictive expression for the eddy viscosity that
provided a remarkable quantitative agreement with experimental data
such as the mean flow rate in turbulent pipe flow (Schlichting 1979).

Despite the success of mixing-length theory, much effort has been
expended in a search for more accurate representations of the Reynolds
stress in turbulent flows, especially in engineering applications.  As
reviewed by Speziale (1991), the more successful approaches start from
the exact equation governing the evolution of the Reynolds stress and
apply a procedure of closure modelling to deal with the numerous
intractable terms that arise.  Through successive algebraic
development accompanied by a large number of parameters, such models
are able to fit an increasing range of experimental or numerical
results.  However, in our view, one disadvantage of this approach in
its current state is a loss of physical interpretation of those terms
derived from complicated algebraic constructions. Moreover, as noted
by Speziale (1991), some of these models tend to perform poorly in
situations for which they were not calibrated, such as rotating shear
flows.

Astrophysics provides examples of naturally occurring shear flows in
which rotation is an essential feature.  Accretion discs (e.g. Pringle
1981) are usually thin discs of gas in circular orbital motion around
a central star or black hole, and are involved in the processes of
star and planet formation as well as being responsible for some of the
most luminous sources in the Universe.  According to Kepler's third
law, the angular velocity of the gas depends on the distance from the
centre as $\Omega\propto r^{-3/2}$, and an outstanding question of
potentially profound significance is whether hydrodynamic turbulence
occurs in these situations (e.g. Balbus \& Hawley 1998).  While it is
true that the Reynolds number is exceedingly large ($Re>10^{14}$;
Frank, King \& Raine 2002), nevertheless the centrifugal instability of
Rayleigh (1917) does not occur because the specific angular momentum
$r^2\Omega$ increases outwards (the same criterion was derived by
Solberg for homentropic compressible fluids; see Tassoul 1978), and
no other suitable hydrodynamic instability has been identified.

Historical experiments by Taylor (1923, 1936{\it a},{\it b}) and Wendt
(1933) on Couette--Taylor flow between differentially rotating
cylinders have been adduced in the hope of a resolution of this issue
(Richard \& Zahn 1999).  The experiments suggest that turbulence can
be sustained even in certain apparently Rayleigh-stable situations
such as a Couette--Taylor flow in which only the outer cylinder
rotates.  In this case the turbulent state is presumably accessed
through a nonlinear shear instability of the laminar state associated
with a subcritical bifurcation, but the wider applicability of this
finding is not well understood. The situation is not helped by the
dearth of modern experiments and the fact that Taylor's findings have
been challenged by Schultz-Grunow (1959).

In this paper we investigate the nonlinear dynamics of turbulent shear
flows, with and without rotation, in the context of a simple but
physically motivated closure of the Reynolds-stress equation.  Our
approach differs from that of the conventional closure models used in
engineering applications. We aim to study a minimal system in which
the modelled nonlinear terms have a clear interpretation and are as
few in number as is compatible with the physical requirements. Indeed,
in astrophysical applications the added complexity of other physical
processes (convection, magnetic fields, radiative transfer, etc.)
forbids anything but a minimal approach in turbulence modelling.  In
the present context of purely hydrodynamic turbulence this approach
allows us to explore the dynamical and nonlinear behaviour in some
detail without losing sight of the physical problem.

The remainder of this paper is organized as follows.  We formulate the
model in Section~2.  Next, in Section~3, we examine the local
properties of the model in the contexts of homogeneous shear
turbulence, with and without rotation, and turbulent shear flow past a
wall.  We also compare the model with others available in the
literature.  
Section~4 concerns the problem of Couette--Taylor flow.  
The results are summarized and discussed in Section~5.

\section{A simple Reynolds-stress model for turbulent shear flows}

\label{sec:model}

The flow of an incompressible fluid is governed by the Navier--Stokes
equations
\begin{equation}
  (\partial_t+u_j\partial_j)u_i=-\partial_ip+\nu\partial_{jj}u_i,
\end{equation}
\begin{equation}
  \partial_iu_i=0,
\end{equation}
where $u_i$ is the velocity, $p$ is the modified pressure (being the
pressure divided by the uniform density $\rho$, plus the gravitational
potential), $\nu$ is the uniform kinematic viscosity, and we make use
of the Cartesian tensor notation.  Following a standard technique, the
velocity and pressure may be separated into mean and fluctuating
parts, e.g.
\begin{equation}
  u_i=\bar u_i+u_i',\qquad
  \langle u_i'\rangle=0,
\end{equation}
where the angle brackets denote a suitable averaging procedure.  We
readily obtain the averaged Navier--Stokes equations,
\begin{equation}
  (\partial_t+\bar u_j\partial_j)\bar u_i=-\partial_i\bar p+
  \nu\partial_{jj}\bar u_i-\partial_jR_{ij},
  \label{dubar}
\end{equation}
\begin{equation}
  \partial_i\bar u_i=0,
  \label{divubar}
\end{equation}
where
\begin{equation}
  R_{ij}=\langle u_i'u_j'\rangle
\end{equation}
is the Reynolds-stress tensor divided by the density. 

From the fluctuating parts of the Navier--Stokes equations it is
possible to obtain an exact equation for $R_{ij}$ in the form
\begin{eqnarray}
  \lefteqn{(\partial_t+\bar u_k\partial_k)R_{ij}+
  R_{ik}\partial_k\bar u_j+R_{jk}\partial_k\bar u_i-
  \nu\partial_{kk}R_{ij}}&\nonumber\\
  &&\qquad=-2\nu\langle(\partial_ku_i')(\partial_ku_j')\rangle-
  \langle u_i'u_k'\partial_ku_j'+u_j'u_k'\partial_ku_i'\rangle-
  \langle u_i'\partial_jp'+u_j'\partial_ip'\rangle.
\end{eqnarray}
There is no difficulty in retaining the exact form of the linear terms
on the left-hand side of this equation, which represent the advection
of the turbulent fluctuations by the mean flow, their interaction with
the mean velocity gradient and the viscous diffusion of the Reynolds
stress.  The difficult terms on the right-hand side cannot be written
exactly in terms of $R_{ij}$, unless further information is known
about the turbulence, and therefore require a closure model.  However,
the physical effects of these terms are quite well understood and this
insight can be used as a guide in constructing the model.  In
particular, the viscous term on the right-hand side is negative
definite and causes a dissipation of the turbulent kinetic energy at a
rate that is usually considered to be independent of $\nu$ in the
limit of large Reynolds number.  The other terms are conservative but
allow for a redistribution of energy among the components of $R_{ij}$,
and it is well established that anisotropic turbulence has a tendency
to return to isotropy (e.g. Rotta 1951).

Recently, one of us proposed a simple model of the stresses in
astrophysical magnetohydrodynamic turbulence (Ogilvie 2003).  In the
special case of hydrodynamic turbulence in an incompressible fluid,
the model reduces to the simpler form
\begin{equation}
  (\partial_t+\bar u_k\partial_k)R_{ij}+R_{ik}\partial_k\bar u_j+
  R_{jk}\partial_k\bar u_i=-{{C_1}\over{L}}R^{1/2}R_{ij}-
  {{C_2}\over{L}}R^{1/2}
  (R_{ij}-{\textstyle{{1}\over{3}}}R\delta_{ij}),
  \label{drij_old}
\end{equation}
where $R=R_{ii}$ is the mean-square turbulent velocity, $C_1$ and
$C_2$ are positive dimensionless constants of order unity, and $L$ is
a characteristic length-scale related to the geometrical constraints
that limit the size of coherent structures.  This model was intended
to represent the typical astrophysical situation in which the Reynolds
number is exceedingly large and there are no solid surfaces on which
boundary layers may form.  The modelled nonlinear terms represent two
well known physical processes that are essential in the dynamics of
turbulent shear flows.  The $C_1$ term represents the viscous
dissipation of turbulent motion, at a rate related to the
characteristic time-scale $L/R^{1/2}$ of the largest eddies.  The
$C_2$ term, which only redistributes energy among the components of
$R_{ij}$, represents the tendency of the turbulence to return to
isotropy on a similar time-scale.  This formulation gives arguably the
simplest nonlinear model involving these two essential effects, which
also guarantees that the Reynolds tensor remains positive definite and
therefore realizable by a genuine velocity field.

In the present paper we are interested in applying the model to
laboratory shear flows in which the Reynolds number is not exceedingly
large and in which boundary layers are present.  We therefore enhance
the model by retaining the viscous diffusion of the Reynolds stress
and including an additional term to model viscous dissipation on the
right-hand side:
\begin{eqnarray}
  \lefteqn{(\partial_t+\bar u_k\partial_k)R_{ij}+
  R_{ik}\partial_k\bar u_j+R_{jk}\partial_k\bar u_i-
  \nu\partial_{kk}R_{ij}}&\nonumber\\
  &&\qquad=-{{C_1}\over{L}}R^{1/2}R_{ij}-
  {{C_2}\over{L}}R^{1/2}
  (R_{ij}-{\textstyle{{1}\over{3}}}R\delta_{ij})-
  {{C_\nu\nu}\over{L^2}}R_{ij}.
  \label{drij}
\end{eqnarray}
Here $C_\nu$ is a third positive dimensionless constant of order
unity, and the $C_\nu$ term allows for the fact that, at low or
moderate Reynolds numbers, when an efficient turbulent cascade does
not form, the viscous dissipation rate is directly proportional to the
viscosity.  Hence in what follows, although we often denote as
`turbulent' any flow for which $R>0$, the Reynolds stresses for
relatively low Reynolds numbers are more likely to represent the
average behaviour of large-scale coherent structures (such as Taylor
vortices in the case of the Couette--Taylor system).  Wavelike
behaviour, on the other hand, cannot be well represented in this
formalism owing to the assumed locality of the dissipation.

In the original model $L$ was related to the thickness of an accretion
disc; in a stratified atmosphere it might be related to the density
scale-height.  It was not necessary to give a precise definition of
$L$ owing to the invariance of the original model under a rescaling
\begin{equation}
  L\mapsto\lambda L,\qquad C_1\mapsto\lambda C_1,\qquad
  C_2\mapsto\lambda C_2.
\end{equation}
In the present paper, however, we will make a definite choice for $L$
appropriate to the geometrical constraints of the problem, and we will
attempt to fix the values of the coefficients by comparison with
experimental results.

It is straightforward, as in Ogilvie (2003), to allow for a uniform
rotation of the frame of reference with angular velocity $\Omega_i$.
In this case we obtain additional Coriolis terms of the form
\begin{equation}
  (\partial_t+\bar u_k\partial_k)R_{ij}+
  R_{ik}\partial_k\bar u_j+R_{jk}\partial_k\bar u_i+
  2\epsilon_{jkl}\Omega_kR_{il}+
  2\epsilon_{ikl}\Omega_kR_{jl}+\cdots.
\end{equation}
The expression for the Reynolds-stress equation (\ref{drij_old}) in a
general orthogonal curvilinear coordinate system is given in
Appendix~A, together with an explicit expression for $\ptl_{kk}R_{ij}$
in cylindrical polar coordinates.

\section{Local properties of the model}
\label{sec:local}

\subsection{Homogeneous shear turbulence}
\label{sec:localhom}

We first consider the idealized situation of a uniform shear flow,
\begin{equation}
  \bar{\bu}=Sy\,\be_x,
\end{equation}
where $(x,y,z)$ are Cartesian coordinates and the shear rate $S$ is
prescribed.  We also allow for a uniform rotation of the frame of
reference with angular velocity
\begin{equation}
  \bOmega=\Omega\,\be_z.
\end{equation}
We assume that the geometrical constraints are such that $L$ is
constant.  This can be achieved by performing a
numerical simulation in a periodic box with no solid boundaries
(Rogallo 1981; Pumir 1996), in which case $L$ is related to the size
of the box. This system is the only one in which the size of the coherent 
structures can be limited without imposing additional boundary conditions 
that would result in the loss of the large-scale homogeneity of the 
flow.  The turbulence may therefore be assumed to be statistically
homogeneous, although it is anisotropic.  In this case the Reynolds
stress depends only on time and the model reduces to a system of
ordinary differential equations constituting an autonomous nonlinear
dynamical system.  We find, in detail,
\begin{eqnarray}
  \partial_tR_{xx}+2(S-2\Omega)R_{xy}&=&
  -{{(C_1+C_2)}\over{L}}R^{1/2}R_{xx}+
  {{C_2}\over{3L}}R^{3/2}-{{C_\nu\nu}\over{L^2}}R_{xx},
  \nonumber\\
  \partial_tR_{xy}+2\Omega R_{xx}+(S-2\Omega)R_{yy}&=&
  -{{(C_1+C_2)}\over{L}}R^{1/2}R_{xy}
  \phantom{{}+{{C_2}\over{3L}}R^{3/2}}-
  {{C_\nu\nu}\over{L^2}}R_{xy},\nonumber\\
  \partial_tR_{xz}+(S-2\Omega)R_{yz}&=&
  -{{(C_1+C_2)}\over{L}}R^{1/2}R_{xz}
  \phantom{{}+{{C_2}\over{3L}}R^{3/2}}-
  {{C_\nu\nu}\over{L^2}}R_{xz},\nonumber\\
  \partial_tR_{yy}+4\Omega R_{xy}&=&
  -{{(C_1+C_2)}\over{L}}R^{1/2}R_{yy}+
  {{C_2}\over{3L}}R^{3/2}-{{C_\nu\nu}\over{L^2}}R_{yy},
  \nonumber\\
  \partial_tR_{yz}+2\Omega R_{xz}&=&
  -{{(C_1+C_2)}\over{L}}R^{1/2}R_{yz}
  \phantom{{}+{{C_2}\over{3L}}R^{3/2}}-
  {{C_\nu\nu}\over{L^2}}R_{yz},
  \nonumber\\
  \partial_tR_{zz}&=&
  -{{(C_1+C_2)}\over{L}}R^{1/2}R_{zz}+
  {{C_2}\over{3L}}R^{3/2}-{{C_\nu\nu}\over{L^2}}R_{zz}.
  \nonumber\\
  \label{dynamical_system}
\end{eqnarray}
It can easily be seen that the two components $R_{xz}$ and $R_{yz}$
are decoupled from the others; these quantities may be expected to
vanish on grounds of symmetry, as is confirmed below.

The system is characterized by a Reynolds number
\begin{equation}
  Re={{L^2|S|}\over{\nu}}
\end{equation}
and an inverse Rossby number
\begin{equation}
  Ro^{-1}={{2\Omega}\over{S}}.
\end{equation}
The Rayleigh discriminant of the rotating shear flow is
\begin{equation}
  \Phi=2\Omega(2\Omega-S).
\label{eq:raydisc}
\end{equation}
We recall that $\Phi<0$ is a sufficient condition for the instability
of a rotating shear flow in the absence of viscosity.  We refer
informally to the situations $\Phi<0$, $\Phi=0$ and $\Phi>0$ as
`Rayleigh-unstable', `Rayleigh-neutral' and `Rayleigh-stable' even
though the criterion is limited in its strict applicability.  It is
convenient to define a dimensionless Rayleigh discriminant
\begin{equation}
  \phi={Ro}^{-1}\left({Ro}^{-1}-1\right)
\end{equation}
and a Taylor number
\begin{equation}
  Ta=-{Re}^2\phi.
\end{equation}

The dynamical system (\ref{dynamical_system}) possesses a trivial
fixed point, $R_{ij}=0$, which represents the laminar state.  Linear
analysis indicates that the trivial fixed point is unstable with
respect to infinitesimal perturbations when the Taylor number exceeds
a positive critical value,
\begin{equation}
  Ta>{Ta}_{\rm c}={{C_\nu^2}\over{4}}.
\end{equation}
Although a linear instability of this kind is possible only in
Rayleigh-unstable situations, we demonstrate below that turbulent
states can be accessed through a nonlinear instability of the laminar
state under a wider range of conditions.

\begin{figure}
  \centerline{\epsfysize12cm\epsfbox{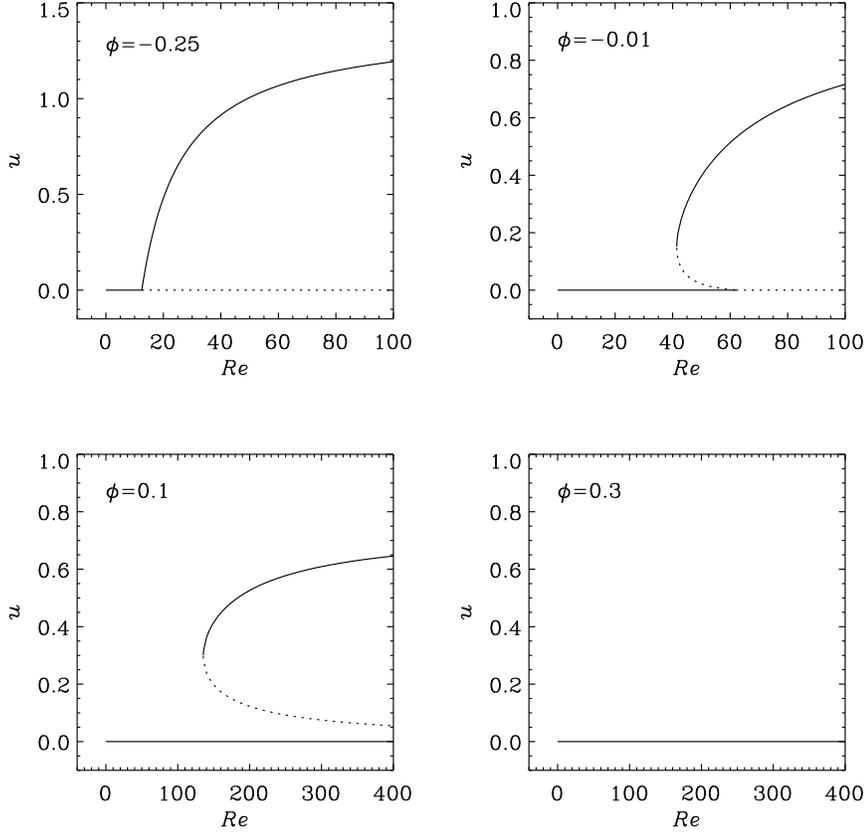}}
  \caption{Bifurcation
    diagrams for homogeneous shear flow with standard model parameters
    ($C_1=0.412$, $C_2=0.6$ and $C_\nu=12.48$) and four different
    values of the dimensionless Rayleigh discriminant $\phi$. For
    these parameter values, $\phi_- \simeq -0.049$ and $\phi_+ \simeq
    0.243$.  Solid and dotted lines indicate stable and unstable
    branches.}
\label{fig:bif1}
\end{figure}

The dynamical system may also possess non-trivial fixed points,
representing states of statistically steady and homogeneous turbulence
in which viscous dissipation is compensated by an extraction of energy
from the shear flow.  Such states may be either stable or unstable;
even if it does not represent a statistical endpoint of the dynamics,
an unstable solution may play a transient role in the dynamics by
providing an organizing structure in the dynamical phase space.
Searching for non-trivial fixed points, we obtain the condition
\begin{equation}
  {{2C_2}\over{3L}}S^2R^{1/2}=
  \left\{\left[{{(C_1+C_2)}\over{L}}R^{1/2}+
  {{C_\nu\nu}\over{L^2}}\right]^2+8\Omega(2\Omega-S)\right\}
  \left[{{C_1}\over{L}}R^{1/2}+{{C_\nu\nu}\over{L^2}}\right],
\end{equation}
which may be written as a cubic equation for the dimensionless rms
turbulent velocity $u=R^{1/2}/L|S|$,
\begin{equation}
  {{2C_2}\over{3}}u=\left\{\left[(C_1+C_2)u+{{C_\nu}\over{Re}}\right]^2+
  4\phi\right\}
  \left(C_1u+{{C_\nu}\over{Re}}\right).
  \label{cubic}
\end{equation}
The behaviour of $u$ as a function of $Re$ depends on the
dimensionless Rayleigh discriminant $\phi$ of the rotating shear flow.
In the present model two values of $\phi$ with special significance
are
\begin{equation}
  \phi_-=-{{C_2}\over{12(C_1+C_2)}},\qquad
  \phi_+={{C_2}\over{6C_1}}.
\end{equation}
Excluding degenerate intermediate cases, we identify four intervals of
interest and illustrate in Figure~\ref{fig:bif1} the corresponding
bifurcation diagrams in which $u$ is plotted against $Re$.  Although
the qualitative features of the set of bifurcation diagrams do not
depend on the parameters $C_1$, $C_2$ and $C_\nu$, we make a
particular selection of `standard' parameters which is explained in
Section~\ref{sec:para} below.
\begin{enumerate}
\item $-{\textstyle{{1}\over{4}}}\le\phi<\phi_-$.  As $Re$ is
  increased, the laminar state loses stability to a branch of
  turbulent solutions at a supercritical bifurcation.
\item $\phi_-<\phi<0$.  The turbulent branch bifurcates
  subcritically from the laminar state at the point of linear
  instability.  There is an interval of $Re$ in which stable laminar
  and turbulent solutions coexist.
\item $0<\phi<\phi_+$.  The laminar state is linearly stable for
  all $Re$ and the turbulent branch is disconnected from it.  Only the
  upper turbulent branch is stable, but the unstable lower branch
  assists in diminishing the basin of attraction of the laminar state
  as $Re$ is increased.
\item $\phi_+<\phi<\infty$.  No turbulent solution exists and the
  laminar state is stable for all $Re$.
\end{enumerate}

\begin{figure}
  \centerline{\epsfysize8cm\epsfbox{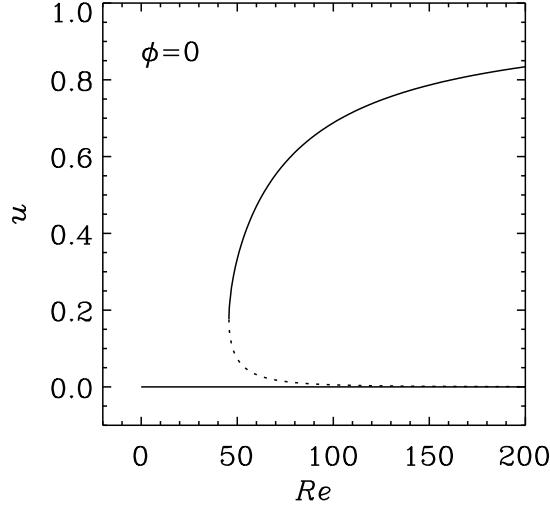}}
  \caption{Bifurcation diagram for homogeneous shear flow with
     standard model parameters and vanishing Rayleigh
    discriminant, corresponding to a non-rotating shear flow.  The
    branch of turbulent solutions bifurcates subcritically from the
    laminar state at $Re=\infty$.}
\label{fig:bif2}
\end{figure}

For non-rotating shear flows, or more generally, 
flows with zero angular-momentum gradient ($\phi=0$), 
the laminar state is linearly stable for all finite values of
$Re$.  However, it is unstable with respect to algebraically growing
disturbances at $Re=\infty$ and a branch of turbulent solutions
bifurcates subcritically at this point (Figure~\ref{fig:bif2}).

The conclusion of this analysis is that, according to our model,
statistically steady and homogeneous turbulence can be sustained in a
rotating uniform shear flow at sufficiently large Reynolds number
provided that the flow is Rayleigh-unstable, Rayleigh-neutral or else
Rayleigh-stable by a sufficiently small margin.  For Rayleigh-neutral
or slightly Rayleigh-stable flows the transition to turbulence occurs
through a nonlinear instability of the laminar state, which has a
diminishing basin of attraction as $Re\to\infty$.  These properties
are in accord with the generally accepted model of transition to
turbulence in shear flows (e.g. Grossmann 2000).  Even though our model
is designed principally to describe the statistical properties of
fully developed turbulence, it appears to give a description of the
onset of turbulence that is at least qualitatively reasonable.

Through straightforward algebra it can be shown that the turbulent
solutions share the following properties:
\begin{enumerate}
\item $R_{ij}$ is positive definite and therefore the solutions
  are realizable;
\item ${\rm sign}(-R_{xy})={\rm sign}(S)$ and therefore the turbulent
  transport of momentum has the same sense as viscous transport;
\item $R_{xz}=R_{yz}=0$ as expected on grounds of symmetry;
\item the solutions are stable with respect to arbitrary
  perturbations of $R_{xz}$ and $R_{yz}$.
\end{enumerate}

In the limit $Re\to\infty$ equation (\ref{cubic}) has at most one
solution for which $u$ tends to a positive limiting value.  This
solution exists when $\phi<\phi_+$, i.e. when the flow is
Rayleigh-unstable, Rayleigh-neutral, or else Rayleigh-stable by a
sufficiently small margin (Ogilvie 2003).  In detail, the limiting
solution is
\begin{eqnarray}
  R_{xx}&=&\left[{{3(1-{Ro}^{-1})C_1+C_2}\over{C_1+C_2}}\right]
  {{R}\over{3}},\nonumber\\
  R_{xy}&=&-{{C_1}\over{2LS}}R^{3/2},\nonumber\\
  R_{yy}&=&\left({{3{Ro}^{-1}C_1+C_2}\over{C_1+C_2}}\right)
  {{R}\over{3}},\nonumber\\
  R_{zz}&=&\left({{C_2}\over{C_1+C_2}}\right){{R}\over{3}},
\end{eqnarray}
with
\begin{equation}
  R=\left[{{C_2-6{Ro}^{-1}({Ro}^{-1}-1)C_1}\over{C_1(C_1+C_2)^2}}\right]
  {{2}\over{3}}L^2S^2=
  {{4(\phi_+-\phi)}\over{(C_1+C_2)^2}}L^2S^2.
\label{eq:r_local}
\end{equation}
For a fixed Rossby number satisfying the condition $\phi<\phi_+$,
  the turbulent momentum transport has the dependence
\begin{equation}
  -R_{xy}={{4C_1}\over{(C_1+C_2)^3}}(\phi_+-\phi)^{3/2}L^2S|S|
  \propto L^2S|S|,
\label{eq:rxy_local}
\end{equation}
as also occurs in Prandtl's mixing-length theory.  In our model,
however, this result is obtained only in the limit of large $Re$, and
the coefficient of proportionality depends on the Rossby number if the
flow is rotating.

The turbulent states are always anisotropic owing to the effects of
shear and rotation.  In our model, the dimensionless anisotropy tensor
\begin{equation}
  b_{ij}={{R_{ij}}\over{R}}-{{1}\over{3}}\delta_{ij},
\end{equation}
which describes the shape of the Reynolds tensor, depends only on the
ratio $C_2/C_1$ and on the Rossby number, when the Reynolds number is
sufficiently large.  When $C_2/C_1$ is small, the tendency to return
to isotropy is weak and the stress becomes highly anisotropic.  In
principle, the ratio $C_2/C_1$ could be constrained through a
comparison with experimental data on anisotropy.  We return to this
point in Section~\ref{sec:para} below.

\subsection{Turbulent shear flow past a wall}
\label{sec:univ_boundlayer}

In this section we analyse the simplest problem involving a
wall-bounded turbulent shear flow.  The solution will serve later as
an asymptotic description of the turbulent boundary layers found in
more complicated situations.

We consider the non-rotating parallel shear flow $\bar{\bu}=\bar
u_x(y)\,\be_x$ in the semi-infinite region $y>0$ bounded by a smooth,
stationary wall at $y=0$ and forced by a shear stress $T_{xy}>0$ at
$y=\infty$.  Even in a situation such as Couette--Taylor flow, it is
usually permissible to neglect rotation in the turbulent boundary
layers because the local shear rate is much larger than the rotation
rate.  Since the presence of the wall provides the only geometrical
constraint on the turbulent structures, we set $L=y$, as is common in
applications of mixing-length theory to wall-bounded flows.  Indeed,
throughout the remainder of this paper, we set $L$ equal to the
distance to the nearest wall.

We seek steady solutions of the averaged equations in which the mean
quantities depend only on $y$, and with the expected symmetry property
$R_{xz}=R_{yz}=0$.  The $x$-component of the averaged Navier--Stokes
equation,
\begin{equation}
  0=\nu\partial_{yy}\bar u_x-\partial_yR_{xy},
\end{equation}
implies
\begin{equation}
  \nu\partial_y\bar u_x-R_{xy}=T_{xy}={\rm constant}.
\label{eq:stat_cartflow1}
\end{equation}
The non-trivial components of the Reynolds-stress equation in our
model are
\begin{eqnarray}
  2R_{xy}\partial_y\bar u_x&=&
  -{{(C_1+C_2)}\over{L}}R^{1/2}R_{xx}+
  {{C_2}\over{3L}}R^{3/2}+
  \nu\partial_{yy}R_{xx}-
  {{C_\nu\nu}\over{L^2}}R_{xx},
  \nonumber\\
  R_{yy}\partial_y\bar u_x&=&
  -{{(C_1+C_2)}\over{L}}R^{1/2}R_{xy}
  \phantom{{}+{{C_2}\over{3L}}R^{3/2}}+
  \nu\partial_{yy}R_{xy}-
  {{C_\nu\nu}\over{L^2}}R_{xy},\nonumber\\
  0&=&
  -{{(C_1+C_2)}\over{L}}R^{1/2}R_{yy}+
  {{C_2}\over{3L}}R^{3/2}+
  \nu\partial_{yy}R_{yy}-{{C_\nu\nu}\over{L^2}}R_{yy},\nonumber\\
  0&=&
  -{{(C_1+C_2)}\over{L}}R^{1/2}R_{zz}+
  {{C_2}\over{3L}}R^{3/2}+
  \nu\partial_{yy}R_{zz}-{{C_\nu\nu}\over{L^2}}R_{zz},
\label{eq:stat_cartflow2}
\end{eqnarray}
subject to the no-slip boundary conditions $\bar u_x=R_{ij}=0$ at
$y=0$.

We rewrite the equations in a dimensionless form by means of the
standard transformations
\begin{equation}
  \bar u_x(y)=v(\eta)\sqrt{T_{xy}},\qquad
  R_{ij}(y)=r_{ij}(\eta)T_{xy},\qquad
  \eta={{y\sqrt{T_{xy}}}\over{\nu}}.
\end{equation}
Observing the property $r_{zz}=r_{yy}$, which implies
$r=r_{xx}+2r_{yy}$, we obtain the reduced problem
\begin{equation}
  v'-r_{xy}=1,
\end{equation}
\begin{eqnarray}
  2r_{xy}v'&=&
  -{{(C_1+C_2)}\over{\eta}}r^{1/2}r_{xx}+
  {{C_2}\over{3\eta}}r^{3/2}+
  r_{xx}''-
  {{C_\nu}\over{\eta^2}}r_{xx},
  \nonumber\\
  r_{yy}v'&=&
  -{{(C_1+C_2)}\over{\eta}}r^{1/2}r_{xy}
  \phantom{{}+{{C_2}\over{3\eta}}r^{3/2}}+
  r_{xy}''-
  {{C_\nu}\over{\eta^2}}r_{xy},\nonumber\\
  0&=&
  -{{(C_1+C_2)}\over{\eta}}r^{1/2}r_{yy}+
  {{C_2}\over{3\eta}}r^{3/2}+
  r_{yy}''-{{C_\nu}\over{\eta^2}}r_{yy},
\end{eqnarray}
with boundary conditions $v(0)=r_{ij}(0)=0$.  

The desired behaviour at $\eta=\infty$ can be deduced by analysing the
far-field limit $\eta\gg1$, for which we obtain the asymptotic form
\begin{eqnarray}
  v'&\sim&v'_1\eta^{-1}+v'_2\eta^{-2}+\cdots,\nonumber\\
  r_{ij}&\sim&r_{ij0}+r_{ij1}\eta^{-1}+\cdots,
\end{eqnarray}
with
\begin{equation}
  v'_1={{C_1}\over{2}}r_0^{3/2},\qquad
  r_0=\left({{6}\over{C_1C_2}}\right)^{1/2}(C_1+C_2),
\end{equation}
\begin{equation}
  r_{xx0}={{(3C_1+C_2)}\over{3(C_1+C_2)}}r_0,\qquad
  r_{xy0}=-1,\qquad
  r_{yy0}={{C_2}\over{3(C_1+C_2)}}r_0.
\end{equation}

\begin{figure}
  \centerline{\epsfysize8cm\epsfbox{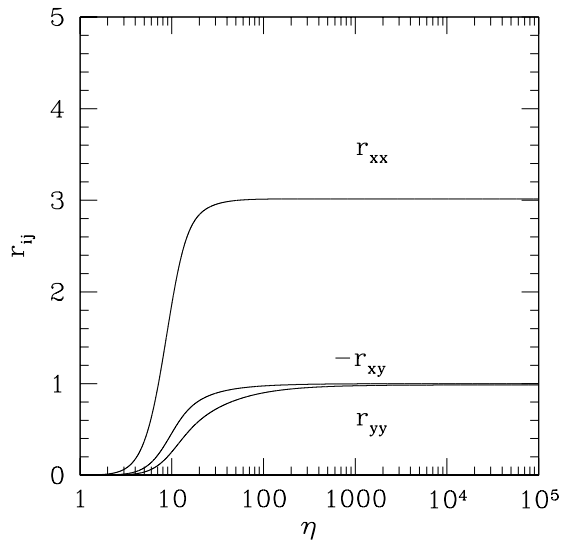}}
  \caption{Universal boundary-layer solution for
    turbulent shear flow past a wall with standard model parameters.
    The dimensionless Reynolds stress components $r_{xx}$,
    $r_{yy}=r_{zz}$ and $-r_{xy}$ are shown.}
\label{fig:univlayer1}
\end{figure}

The reduced problem is universal, involving no parameters other than
the model parameters $C_1$, $C_2$ and $C_\nu$.  To solve it
numerically, we use a finite-difference Newton--Raphson relaxation
method on a stretched mesh, starting from an initial guess with
$r_{ij} = r_{ij0}$ and $v = \eta$.  The outer boundary condition
$r_{ij}'(\eta_{\rm out})=0$, where $\eta_{\rm out}\gg1$, imposes the
desired far-field behaviour with adequate fidelity.  We choose
$\eta_{\rm out} = 10^5$.  The desired universal boundary-layer
solution is shown in Figure~\ref{fig:univlayer1}. The stability of
this solution has been confirmed using a time-dependent numerical
method.

In the Prandtl--von K\'arm\'an analysis of turbulent boundary layers
(e.g. Schlichting 1979) the velocity profile for $\eta\gg1$ is given
as
\begin{equation}
  v\simeq A\ln\eta+B
\end{equation}
where $A$ and $B$ are dimensionless empirical constants, and it is
customary to refer to $\kappa=1/A$ as the von K\'arm\'an constant.
This `universal velocity profile', `log law' or `law of the wall' is
generally in excellent agreement with experimental data.  The values
traditionally assigned on the basis of Nikuradse's experiments in the
1930s are $\kappa=0.41$ and $B=5.2$ (for a smooth wall).  However,
recent experiments at higher Reynolds numbers show a much superior fit
with $\kappa=0.436$ and $B=6.15$ (Zagarola \& Smits 1998).

\begin{figure}
  \centerline{\epsfysize8cm\epsfbox{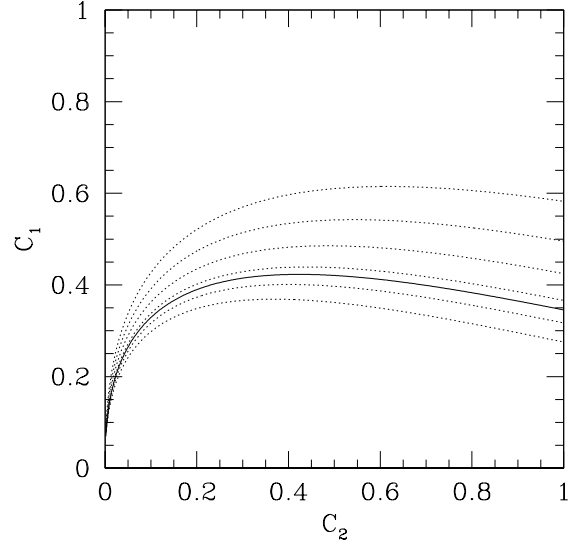}}
  \caption{Contours of constant $\kappa$ with (from bottom to top) 
    $\kappa = 0.5$, $0.46$, $0.436$ (solid line), $0.42$, $0.38$,
    $0.34$ and $0.3$.}
\label{fig:karmann}
\end{figure}

In our model the integrated velocity profile for $\eta \gg 1$ is
\begin{equation}
   v(\eta) = v_0 + v_1' \ln\eta - \frac{v_2'}{\eta} + O(\eta^{-2}),
\label{eq:veta}
\end{equation}
which is asymptotically equivalent to the Prandtl--von K\'arm\'an
velocity profile; thus we identify the von K\'arm\'an constant as
\begin{equation}
  \kappa = \frac{1}{v_1'} = \frac{2}{C_1}\left[\frac{C_1C_2}{6(C_1+C_2)^2}\right]^{3/4}.
\label{eq:vk}
\end{equation}
The von K\'arm\'an constant depends only on $C_1$ and $C_2$ and can
therefore be fitted independently of $C_\nu$.
Figure~\ref{fig:karmann} shows the relation between $C_1$ and $C_2$
for constant $\kappa$.

\begin{figure}
  \centerline{\epsfysize8cm\epsfbox{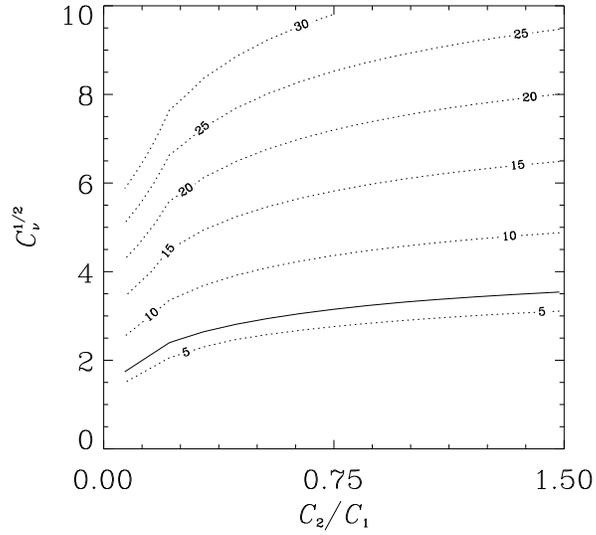}}
  \caption{Contours of constant $v_0$ as a function of $C_2/C_1$ (for a 
    fixed $\kappa = 0.436$) and $\sqrt{C_\nu}$. The solid line
    corresponds to the experimentally determined value of $v_0=6.15$.
    The uniform contour spacings suggest that $v_0$ depends primarily
    and approximately linearly on $\sqrt{C_{\nu}}$.}
  \label{fig:v0}
\end{figure}

The additive constant $B=v_0$ cannot be deduced from the asymptotic
analysis but must instead be determined from a numerical solution of
the problem including the viscous sublayer close to the wall.
Numerical integration of the boundary layer equations shows that $v_0$
depends primarily on $C_\nu$, as shown in Figure~\ref{fig:v0}.  For
simplicity, in that figure we have reduced the parameter space by
imposing the constraint $\kappa = 0.436$ (Zagarola \& Smits 1998).

\subsection{Comparison with other models}

In comparison with other Reynolds-stress models available in the
literature, our model appears quite simplistic (e.g. Speziale 1991).
The viscous dissipation rate is given by
\begin{equation}
  \epsilon={{C_1}\over{2L}}R^{3/2}+{{C_\nu\nu}\over{2L^2}}R
\end{equation}
and we do not attempt to model a separate time-dependent equation for
this quantity.  One reason for this is that the length-scale $L$ is
imposed by the geometrical constraints in our problem, and is not free
to expand as occurs when turbulence is generated in a localized region
within a larger system.  Instead, we allow for the effects of a finite
Reynolds number by specifying a dissipation time-scale that is related
to the characteristic time-scale $L/R^{1/2}$ of the largest eddies in
high-Reynolds-number turbulence, and to the viscous time-scale
$L^2/\nu$ at lower Reynolds numbers.

In freely decaying turbulence with no mean shear or rotation, the
return to isotropy is described in our model by the equation
\begin{equation}
\frac{\dd b_{ij}}{\dd \tau} = \frac{R}{\epsilon} \frac{\dd b_{ij}}{\dd t} =
 - \left(\frac{2C_2}{C_1+C_\nu\nu L^{-1}R^{-1/2}}\right) b_{ij},
\end{equation}
where
\begin{equation}
b_{ij} = \frac{R_{ij}}{R} - \frac{1}{3} \delta_{ij}
\end{equation}
is the anisotropy tensor and $\tau$ is a dimensionless time variable
(e.g. Speziale 1991).  In the limit of large Reynolds number we
therefore obtain a linear return to isotropy, equivalent to the one
introduced first by Rotta (1951).

Furthermore, we do not attempt to model the `rapid pressure--strain
correlation', for which elaborate algebraic models have been proposed
(e.g. Sj\"ogren \& Johansson 2000).  Through this simplification we
may lose some accuracy.  However, it is the treatment of this term
that has given rise to models that deal poorly with rotating shear
flows.  For example, the widely adopted model of Launder, Reece \&
Rodi (1975) is not consistent with Rayleigh's criterion in the sense
that it does not permit turbulence to be sustained at high Reynolds
number in certain situations where Rayleigh's stability criterion is
not satisfied (Speziale 1991), perhaps because a realizability
condition of some kind is implicitly violated.  It is to be hoped that
a way can be found to model the rapid pressure--strain correlation in
future with due regard to Rayleigh's criterion.

\subsection{Summary of the constraints on the parameters $C_1$, $C_2$ and
  $C_{\nu}$}
\label{sec:para}

\begin{figure}
  \centerline{\epsfysize8cm\epsfbox{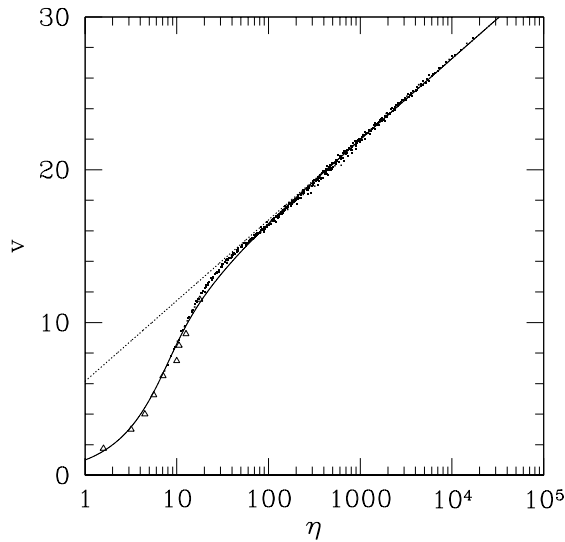}}
  \caption{Solution of the dimensionless boundary-layer problem for
    turbulent shear flow past a wall with standard model parameters.
    The dotted line corresponds to the asymptotic profile $v =
    (0.436)^{-1} \ln \eta + 0.615$ (Zagarola \& Smits 1998) which is
    their suggested best fit for their experimental results. The small
    black dots are the experimental data points from Zagarola \&
    Smits, corrected for systematic errors according to the prescriptions
    of McKeon et al. (2003). The open triangles are the experimental 
    data points from Reichardt (1940).}
\label{fig:zs98}
\end{figure}

We have purposely adopted a simple closure model for the
Reynolds-stress equation so that we can focus on the nonlinear
dynamical properties of the system rather than engaging in a lengthy
exercise of parameter fitting.  Nevertheless, because the model
naturally predicts a logarithmic velocity profile close to a wall, it
makes sense to apply the two accurate constraints $\kappa = 0.436$ and
$v_0 = 6.15$ provided by the very high-quality experimental data on
wall-bounded turbulent shear flows in the Superpipe experiment
(Zagarola \& Smits 1998).  The first constraint provides a relation
between $C_1$ and $C_2$ only, whereas the second (which applies only
for a smooth wall) yields $C_\nu$ provided that $C_1$ and $C_2$ are
known. A third constraint, which would be required to fix all three
parameters of our model, might in principle be provided by
experimental data on the anisotropy of the Reynolds stress in
homogeneous shear turbulence, or on the return to isotropy of
homogeneous turbulence.  In fact, the limitations of the
three-parameter model mean that no choice of the parameters can
accurately match all experimental results.  For example, Choi \&
Lumley (2001) find that the return to isotropy of homogeneous
turbulence is more complicated than is assumed in any available
closure model.  In Section~\ref{sec:datacyl} below we compare the
predictions of our model with data from Couette--Taylor experiments,
and tentatively deduce an approximate value of $C_2 \simeq 0.6$. Hence
in what follows (unless otherwise mentioned) we shall take as standard
parameters $C_1 = 0.412$, $C_2 = 0.6$ and $C_\nu = 12.48$, which yield
$\kappa = 0.436$ and $v_0= 6.15$ as required.  The predicted
boundary-layer velocity profile with this choice of parameters is
compared with the experimental measurements of Zagarola \& Smits
(1998) in Figure~\ref{fig:zs98}.

\section{Couette--Taylor flow}
\label{sec:ctf}

Couette--Taylor flow between differentially rotating coaxial cylinders
is a seemingly simple dynamical system that has been found to exhibit
a rich variety of nonlinear behaviour.  Much of this interesting
dynamics occurs close to the onset of Rayleigh's centrifugal
instability, albeit in a confined setting and in the presence of
viscosity.  Our main interest here is in the existence and properties
of turbulent states in Couette--Taylor flow at large Reynolds numbers,
rather than the behaviour close to the onset of instability.  This
aspect of the problem has received rather little attention from
experimentalists or turbulence modellers.  Recently, however,
arguments based on the Couette--Taylor system have been made in
connection with important questions relating to turbulence in
astrophysical flows involving differential rotation (e.g. Richard \&
Zahn 1999).
  
Since our Reynolds-stress model is based on a covariant formulation,
it naturally includes the effect of rotation as well as shear.  In
this section we apply the model to the Couette--Taylor system, compare
its predictions with the available experimental results and examine
the wider consequences of these findings.

\subsection{Predictions of the model}

\subsubsection{Governing equations and numerical solution}

We consider the shear flow between two infinite coaxial cylinders
located at radii $r_i$ and $r_o$, rotating with angular velocities
$\Omega_i$ and $\Omega_o$.  Adopting a cylindrical coordinate system
$(r,\phi,z)$, we seek steady solutions of the averaged equations in
which the mean quantities depend only on $r$ and the mean flow is
azimuthal only: $\bar{\bu}=r\Omega(r)\,\be_\phi$.  In this case the
Reynolds-stress equation in our model reduces to (\ref{eq:Reqcyl}) and
straightforward algebra yields $R_{rz} = R_{\phi z} = 0$.  We choose
the scale-length $L$ to be the distance to the nearest wall, namely $L
= \min (r-r_i,r_o-r)$.

Experiments on Couette--Taylor flow, using a cylindrical container of
finite height $h$, are perturbed by end-effects, in
particular the Ekman circulation, when the aspect ratio
$h/(r_o-r_i)$ is not very large.  We do not attempt to model
end-effects, but note their potential significance when comparing our
findings with experimental results (see Section~\ref{sec:discdata}).

The angular velocity profile $\Omega(r)$ between the cylinders is
obtained self-consistently by solving also the angular momentum
conservation equation (the azimuthal component of the averaged
Navier--Stokes equation),
 \begin{equation}
\frac{{\rm d}}{ {\rm d}r} \left(r^2 R_{r\phi} +  \nu r^2 S\right) = 0
\label{eq:Amomcyl}
\end{equation}
where $S=-r\,{\rm d}\Omega/{\rm d}r$ is the shear rate in cylindrical
geometry.  Note that equation (\ref{eq:Amomcyl}) can be integrated to
introduce the torque $T$ between the cylinders,
\begin{equation}
r^2 R_{r\phi} +  \nu r^2 S = \frac{T}{2\pi h \rho}.
\label{eq:am}
\end{equation}
We obtain a ninth-order system of ODEs with one eigenvalue ($T$) which
requires ten boundary conditions: $R_{rr} = R_{r\phi} = R_{\phi\phi} =
R_{zz} = 0$ on each boundary, as well as $\Omega(r_i)=\Omega_i$ and
$\Omega(r_o) = \Omega_o$. We solve this two-point boundary-value
problem numerically using a Newton--Raphson relaxation method, using
the solution for $Re \rightarrow \infty$ as an initial guess (or,
whenever applicable, the results of a previous calculation for similar
parameters).

In addition to possible turbulent solutions, there is of course the
well-known laminar Couette--Taylor flow $\Omega(r) = \alpha +
\beta/r^2$, where $\alpha = (\Omega_o r_o^2 - \Omega_i
r_i^2)(r_o^2-r_i^2)^{-1}$ and $\beta = (\Omega_i - \Omega_o)
r_i^2r_o^2 (r_o^2-r_i^2)^{-1}$.  The local dimensionless Rayleigh
discriminant of the laminar solution is $\phi(r) = (\alpha/\beta)^2
r^4 + (\alpha/\beta) r^2$.  Unlike plane Couette flow, the laminar
Couette--Taylor flow is linearly unstable in certain regions of the
parameter space.  Therefore turbulent states may be accessed through
either linear or nonlinear instabilities.

\subsubsection{Asymptotic analysis}
\label{sec:asympcyl}

The results presented in Section~\ref{sec:univ_boundlayer} suggest
that the system of equations (\ref{eq:Reqcyl}) and (\ref{eq:am}) could
also be solved approximately by asymptotic matching, between the
universal boundary-layer solution near the wall and a high-Reynolds
number limiting solution ($Re \rightarrow \infty$) in the main body of
the fluid. When $Re \rightarrow \infty$, the turbulent solution is
\begin{equation}
R =  \left[ \frac{C_2 - 6 Ro^{-1}(Ro^{-1}-1)C_1}{C_1 (C_1+C_2)^2} \right]\frac{2}{3} L^2 S^2,
\end{equation}
whenever this is positive, and then
\begin{equation}
R_{r\phi} = \frac{C_1}{2LS} R^{3/2} = \frac{C_1}{2} \left(\frac{2}{3}\right)^{3/2} \left[ \frac{C_2 - 6C_1 Ro^{-1} ( Ro^{-1} -1)}{C_1(C_1+C_2)^2}\right]^{3/2} L^2 S|S|,
\end{equation}
by direct analogy with results (\ref{eq:r_local}) and
(\ref{eq:rxy_local}) of the local analysis.  Unfortunately, there
exist no analytical solutions to the angular momentum equation with
this Reynolds stress prescription unless $Ro \gg 1$. The Rossby number
of the laminar flow is large for typical narrow-gap setups; one might
therefore hope to use this asymptotic limit for the study of the
turbulent regime also. However, numerical solution of equations
(\ref{eq:Reqcyl}) and (\ref{eq:am}) reveals that turbulence
effectively reduces the shear outside the boundary layers and prevents
the use of the $Ro\gg 1$ asymptotic limit unless the gap is extremely
small (typically, less than a few percent of the average radius). For
completeness, we nevertheless provide such an asymptotic analysis in
Appendix B.

\subsubsection{Stability diagram and structure of solutions across 
parameter space}
\label{sec:stabsols}

\begin{figure}
   \centerline{\epsfysize8cm\epsfbox{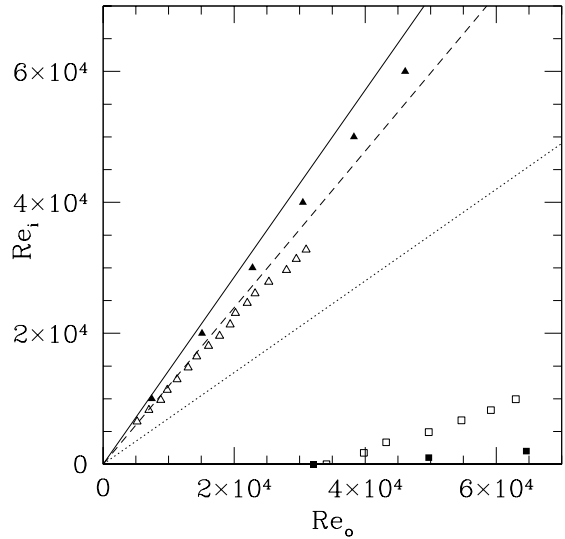}}
   \caption{Stability boundaries for Couette--Taylor flow in a fixed
     geometrical setup ($r_i/r_o=0.7$), predicted with standard model
     parameters. The black symbols are the predictions of our model,
     and delimit the regions of turbulent solutions in the top-left
     and bottom-right corners. The open symbols are the data from
     Richard (2001) for the same geometrical setup. The solid line is
     the stability limit according to Rayleigh's criterion ($\Omega_i
     r_i^2 = \Omega_o r_o^2$), the dashed line marks the Keplerian
     ratio ($\Omega_i r_i^{3/2} = \Omega_o r_o^{3/2}$) and the dotted
     line marks solid-body rotation ($\Omega_i = \Omega_o$).}
\label{fig:stabbound}
\end{figure}

In what follows, we call the Couette--Taylor flow `unstable' whenever
there exist solutions to the equations of the model with $R>0$. By
doing so, we implicitly assume that the background noise level is
sufficiently high to excite finite-amplitude instabilities if they
exist. Figure~\ref{fig:stabbound} shows a stability diagram for
corotating cylinders in the $(Re_o, Re_i)$ plane (where $Re_o = dr_o
\Omega_o/\nu$ and $Re_i = d r_i \Omega_i /\nu$, with $d = r_o-r_i$) for
a given geometrical setup ($r_i/r_o=0.7$).
As predicted by the local analysis, Rayleigh-stable flows can be
subject to finite-amplitude instabilities provided they are notionally
Rayleigh-stable by a small margin only, thus displacing the stability
boundary in the top-left corner of the diagram by a small amount.
Strong enough shear can overcome the stabilizing angular-momentum
gradient in the case where $Re_o \gg Re_i$ and finite-amplitude
instabilities are also found in that region (bottom-right corner of
the diagram). The predicted onset of instability in the case of
relatively low Reynolds numbers and also of counter-rotating cylinders
is discussed further in Section~\ref{sec:onset}. Comparison with
experimental data from Richard et al. (2001) shows significant
discrepancies with the predictions of the model,
though this could be expected because the aspect ratio of 
their experimental setup is not large (see the discussion in
Section~\ref{sec:discdata}).

\begin{figure}
   \centerline{\epsfysize14cm\epsfbox{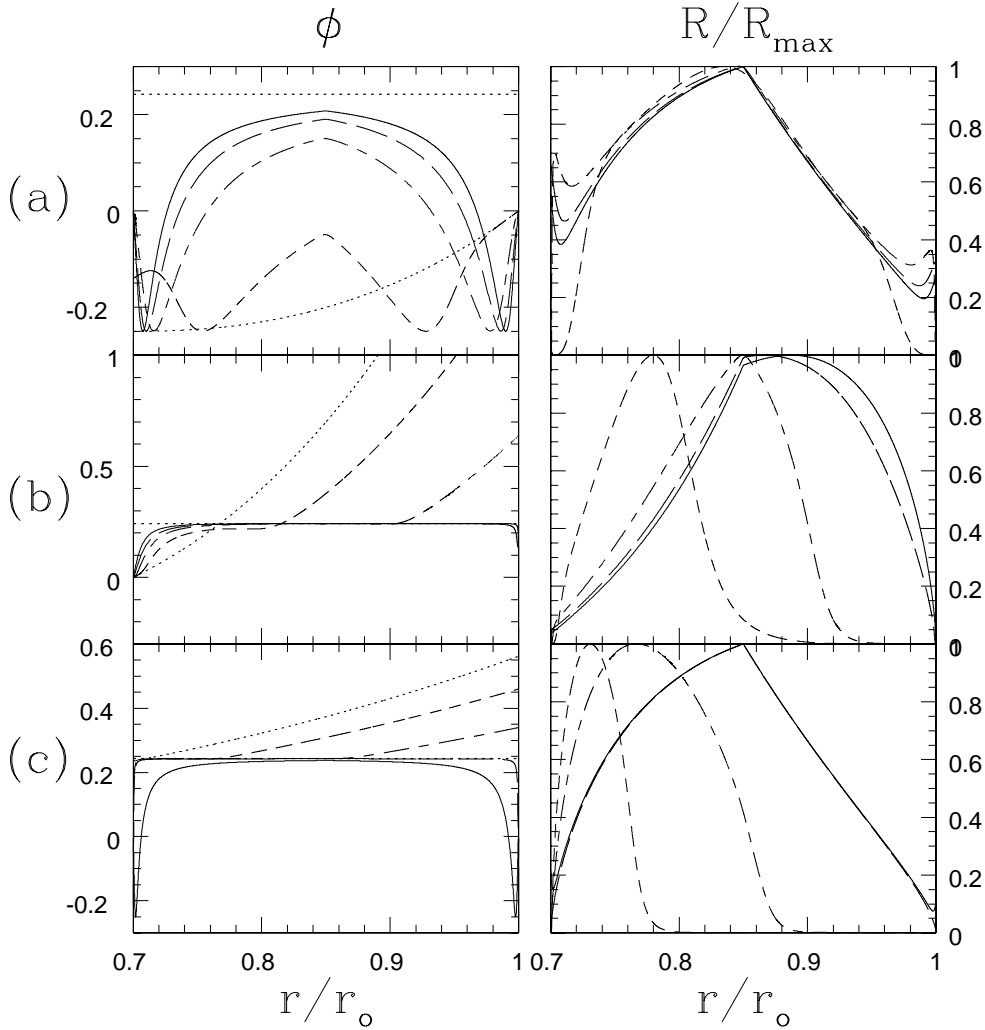}} 
   \caption{Structure of
     solutions for a fixed geometrical setup with $r_i/r_o=0.7$, for
     three regions of parameter space in the case of corotating
     cylinders. The left-hand column shows the local dimensionless
     Rayleigh discriminant $\phi(r)$; in each figure, the horizontal
     dotted line marks the position of the critical value $\phi_+ =
     C_2/6C_1 \simeq 0.243$. The right-hand column shows the
     mean-square turbulent velocity, normalized to its maximum value
     for clarity. The top two panels (a) correspond to the case where
     the outer cylinder is at rest, with the laminar solution marked
     as a dotted line, then the outer cylinder rotation rate is
     steadily increased with $Re_o = 10^3$ (dashed line), $Re_o =
     10^5$ (short-long dashed line), $Re_o =10^7$ (long-dashed line)
     and $Re_o = 10^9$ (solid line).  The line-style coding is the
     same for both plots.  Panels (b) correspond to the case when the
     inner cylinder is at rest, showing the laminar solution (dotted
     line), then turbulent solutions for $Re_o = 10^5$ (dashed line),
     $Re_o=10^6$ (short-long-dashed line), $Re_o = 10^7$ (long-dashed
     line) and $Re_o = 10^9$ (solid line).  Panels (c) correspond to
     the onset of instability near the Rayleigh stability limit. The
     outer cylinder rotation is fixed ($Re_o = 10^9$) and the inner
     cylinder rotation is varied. The solution is laminar (dotted
     line) for $Re_i=1.22 \times 10^9$, then becomes successively more
     turbulent for $Re_i = 1.24 \times 10^9$ (dashed line), $Re_i =
     1.26 \times 10^9$ (short-long-dashed line) and $Re_i = 1.28
     \times 10^9$ (long-dashed line).  The last curve lies further
     away from onset, and into the Rayleigh-unstable zone with $Re_i =
     2 \times 10^9$ (solid line).}
\label{fig:sols}
\end{figure}

The structure of the solutions in various regions of parameter space,
as shown in Figure~\ref{fig:sols}, reflects the physics of turbulent
flow.
In the case where the outer cylinder is at rest, the local
dimensionless Rayleigh discriminant for the laminar solution is
\begin{equation}
\phi(r) = (r/r_o)^4 - (r/r_o)^2
\end{equation}
and is therefore always negative (Rayleigh-unstable). Viscosity
stabilizes the flow for sufficiently low Reynolds number $Re_i$ but
the transition to turbulence occurs directly, through a linear
instability (the critical value for the transition depends on the
geometrical setup). Turbulent stresses are largest in the bulk of the
fluid, near the mid-point $r_m =
(r_o+r_i)/2$.

In the case where the inner cylinder is at rest,
\begin{equation}
\phi(r) = (r/r_i)^4 - (r/r_i)^2
\end{equation}
is always positive (Rayleigh-stable).  Linear instability is therefore
not expected, but turbulent states may still be accessed through a
nonlinear instability.  Sufficiently close to the inner cylinder, in
the region where $\phi<\phi_+$, the local analysis would suggest that
the laminar solution is unstable to finite-amplitude perturbations
provided the local Reynolds number is large enough.  Depending on the
gap width, two situations may arise: either $0 < \phi < \phi_+$ for
all $r$, or there exists a transition within the fluid between a
locally stable region and a region of finite-amplitude instability.
An equivalent way of looking at the problem is to note that the
stabilizing effect of rotation on the development of turbulence is
weaker near the inner boundary when the inner cylinder is at rest, and
so we expect turbulence to develop first near the inner cylinder (as
can indeed be seen in Figure~\ref{fig:sols}).

Finally, we explore the behaviour of solutions near the onset of
nonlinear instability at very high Reynolds number in the region close
to the Rayleigh line ($\Omega_ir_i^2=\Omega_or_o^2$).  The transition
to turbulence occurs when $\phi$ for the laminar solution near the
inner cylinder drops below $\phi_+$.  This happens in the
Rayleigh-stable domain.  As $Re_i$ is increased, $\phi$ on the outer
cylinder drops below $0$ which marks the transition to the
Rayleigh-unstable domain.
 
The structure of the solutions in all three cases seems to suggest
that turbulence is extremely efficient in transporting the applied
torque: the marginal solution $\phi = \phi_+$ is favoured near onset
and also in the Rayleigh-stable case far from onset. In the latter
case the solution deviates from the marginal stability solution only
in the thin viscous boundary layers. This behaviour is typically also
observed in convection.  In the Rayleigh-unstable case on the other
hand, a solution with $\phi = \phi_+>0 $ could not possibly satisfy
the applied boundary conditions, and the flow appears to compromise by
choosing an intermediate solution with $\phi<0$ in a significant part
of the domain.



\subsection{Comparison with experimental data}
\label{sec:datacyl}

\subsubsection{Discussion of available data}
\label{sec:discdata}

Since Taylor's (1923) pioneering work on the stability of fluid flows
between two coaxial rotating cylinders, a wealth of experimental data
has been collected on the dynamical properties of such flows for
various aspect parameters ($r_i,r_o,h$) and for a very large region of
the ($Re_o,Re_i$) parameter space. In particular, Wendt (1933) and
Taylor (1936{\it a},{\it b}) presented the most extensive collection
of torques and velocity measurements for turbulent Couette--Taylor
flow far from onset, whereas Andereck, Liu \& Swinney (1986) reviewed
the successive flow-pattern transitions near onset.

Amongst other notable results, Wendt (1933) studied the contaminating
Ekman flows arising from end-effects, which can drive significant
deviations from the laminar Couette--Taylor flow.  He proposed an
ingenious system including a free top surface and a differentially
rotating split bottom plate to reduce end-effects; this setup is
indeed able to reduce meridional flows but not completely.
Comparisons of torque measurements between various bottom boundary
conditions revealed that end-effects are especially important for
aspect ratios $h/d$ smaller than $40$.  Wendt performed experiments
with only the outer cylinder rotating, and showed that torque
measurements made with bottom plates corotating with the outer
cylinder were roughly 10\% larger than in the case where the bottom
plates are stationary for an aspect ratio of 50, 100\% larger for an
aspect ratio 23 and 400\% larger for an aspect ratio of 11. Naturally,
any theory that assumes axial translational symmetry for the system
(as does the model investigated here) can only be compared with
experiments that have little contamination from end-effects, and we
may use Wendt's findings as a guideline for distinguishing between
adequate and inadequate sets of experiments.

Experiments at very high Reynolds number, and with a wide gap
($r_i/r_o = 0.724$), have been performed in the case where only the
inner cylinder is rotating by Lathrop, Fineberg \& Swinney (1992) and
more recently by Lewis \& Swinney (1999). Richard et al. (2001) have
performed such experiments ($r_i/r_o = 0.7$) with both inner and outer
cylinders rotating, using split-bottom boundary conditions.  Wide-gap
setups are well suited to verify the adequacy of our theory in
describing the effects of rotation on turbulent shear flows (as
opposed to narrow-gap setups, which are principally dominated by the
shear instability).  However, all these experiments have too small
($\le 25$) an aspect ratio to be free of end-effects, so we cannot use
them reliably for comparison with our work.

In what follows (Section~\ref{sec:torquecomp}), we first compare the
predictions of our model with the torque measurements from Wendt
(1933) in order to obtain a third constraint on our parameters $C_1,
C_2$ and $C_\nu$, as anticipated in Section~\ref{sec:para}. We then
compare the model to Taylor's (1936{\it a}) data. In
Section~\ref{sec:onset} we look at the predictions of the model for
the onset of linear instability in Couette--Taylor flows and compare
it with Taylor's (1923) experimental work. Finally, we discuss the
stability of Keplerian shear flows in the light of our model, and
compare our predictions with those of Richard \& Zahn (1999) in
Section~\ref{sec:kep}.

\subsubsection{Torque measurements}
\label{sec:torquecomp}

\begin{figure}
  \centerline{\epsfysize8cm\epsfbox{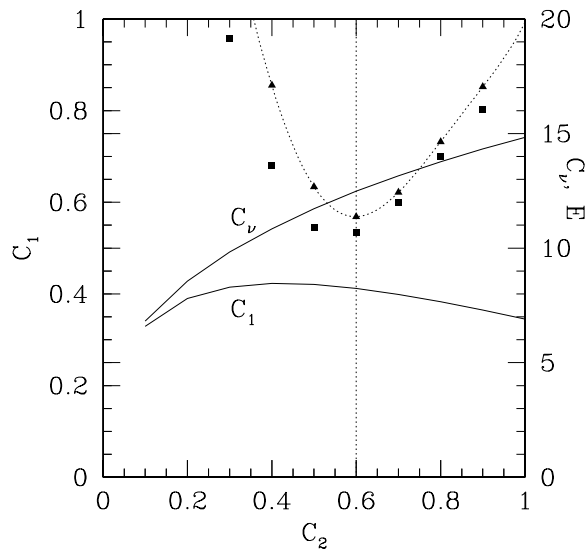}}
  \caption{Study of the
    model parameters that best fit Wendt's (1933) experimental data.
    For each value of $C_2$, $C_1$ is chosen such that $\kappa =
    0.436$ and $C_\nu$ is chosen such that $v_0= 6.15$. The
    corresponding values of $C_1$ and $C_\nu$ are shown as solid
    lines, with the relevant scales on the left and right of the plot
    respectively. Using these parameters, the error $E$ (as given by
    equation \ref{eq:E}) is calculated and shown as triangular
    symbols. The minimum error occurs for $C_2 = 0.6$. We also plot
    the error $E$ calculated when leaving out the experimental
    results for stationary inner cylinder as square symbols.}
\label{fig:parafit}
\end{figure}

\begin{figure}
   \centerline{\epsfysize8cm\epsfbox{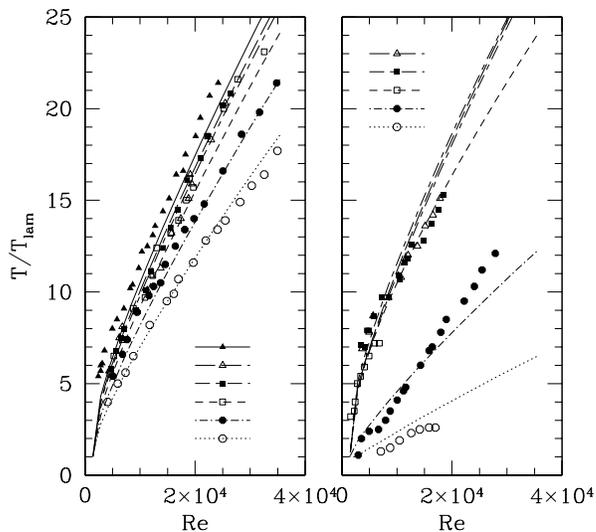}}
   \caption{Comparison between the predictions of the model for $C_1 =
     0.412$, $C_2 = 0.6$ and $C_\nu = 12.48$ and Wendt's (1933) data
     extracted from his article, Fig. 9c. The left panel corresponds
     to his measurements for counter-rotating cylinders, and the right
     panel corresponds to corotating cylinders. The corresponding
     line-styles and symbols are shown in each diagram. }
\label{fig:wendt}
\end{figure}

Wendt's (1933) torque measurements present the most extensive results
for an experimental setup with large aspect ratio. By using his
narrow-gap data, which are relatively free from contaminating
end-effects, we attempt to constrain our basic parameters further.
Wendt presents the results of twelve sets of experiments for the
following setup: $h = 50$ cm, $r_o = 14.70$ cm and $r_i = 13.75$ cm.
For each set of measurements, the ratio of the rotation rates of the
inner and outer cylinders is fixed, and the Reynolds
number\footnote{Wendt actually uses a quantity related to the Reynolds
  number, $(60/2\pi)|\Omega_o-\Omega_i|/\nu$.}  is defined as
\begin{equation}
Re = \frac{|\Omega_o-\Omega_i| r_m d}{\nu} \mbox{   .}
\label{eq:Reboth}
\end{equation}
Wendt plots in his Figure~9c the ratio of the turbulent to the laminar
torque. We have performed ten suites of numerical calculations where
$C_2$ is varied between 0.1 and 1 by increments of 0.1, and for each
set have calculated the typical error between the predictions of our
model and Wendt's experimental data points with the formula
\begin{equation}
E(C_2) = \sum \left[ \left. \frac{T}{T_{\rm lam}} \right|_{\rm mod} -
\left. \frac{T}{T_{\rm lam}} \right|_{\rm exp} \right]^2 \mbox{   ,}
\label{eq:E}
\end{equation}
where the sum spans all data points in all twelve sets of experiments.
The results are shown in Figure~\ref{fig:parafit} and suggest that the
best fit is obtained with parameters $C_1 = 0.412$, $C_2 = 0.6$ and
$C_\nu = 12.48$, which we have adopted as standard throughout this
paper.
The corresponding fit to Wendt's experimental data is shown in
Figure~\ref{fig:wendt}. We note that the fit is quite good though we
are not able to fit all the curves equally well. In particular, the
experimental results for a stationary inner cylinder (black circles on
the right-hand panel) seem to deviate significantly from the
predictions of our model for all possible values of $C_2$.
Leaving this particular set of experiments out of the least-square
fitting procedure seems to reduce the optimal value of $C_2$, but not
significantly (see the square symbols in Figure~\ref{fig:parafit},
which have a minimum near $C_2 = 0.55$). We emphasize that the
constraint on our parameters obtained by fitting Wendt's data is much
less satisfactory than the two tight constraints provided by the
universal velocity profile of turbulent boundary layers.  Therefore
the estimate $C_2=0.6$ is to be regarded as tentative and approximate
only.

\begin{figure}
   \centerline{\epsfysize14cm\epsfbox{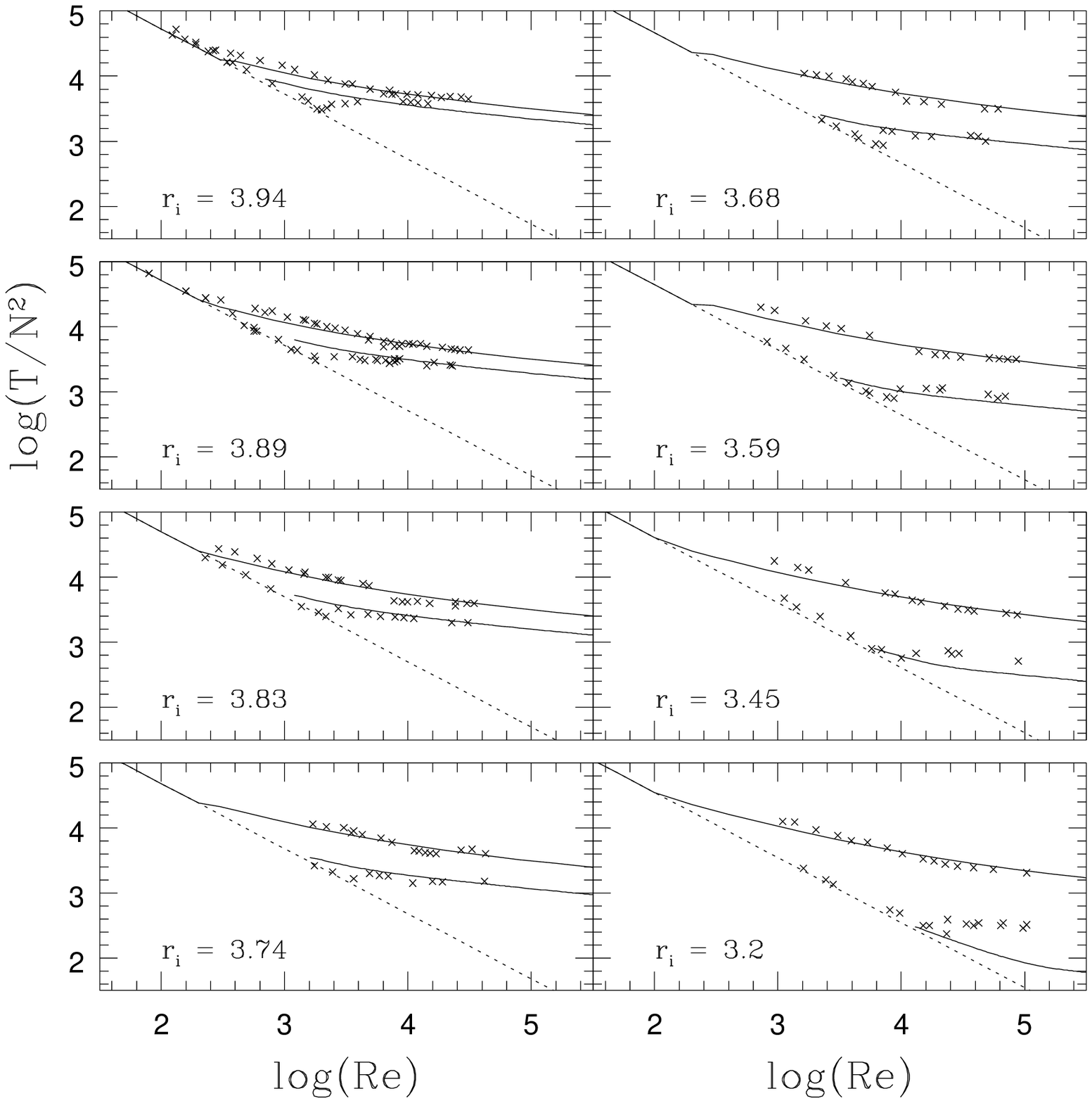}}
   \caption{Comparison between Taylor's original data (1936{\it a}) and the
     predictions of our model with standard parameters.  For each
     panel, the outer cylinder has radius $r_o = 4.05$ cm and the
     inner cylinder radius $r_i$ is given in cm. The upper branch
     corresponds to the torques measured on the outer cylinder when
     the inner cylinder is rotating, and the lower branch corresponds
     to torques measured on the inner cylinder when the outer cylinder
     is rotating. The angular rotation rate $\Omega$ is related to $N$
     as $\Omega = 2\pi N$. The dotted line shows the laminar solution
     whereas the solid line is the prediction of the model. The
     quantity $T/N^2$ is measured in CGS units.}
\label{fig:torque8}
\end{figure}

We then tested the predictions of the model against Taylor's (1936{\it
  a}) torque measurements.  Taylor's data consists of eight sets of
experiments for varying gap width; he compares the torques measured
for similar Reynolds numbers\footnote{Taylor actually uses a quantity
  related to the Reynolds number, $\Omega/2\pi\nu$.}  (as defined by
equation (\ref{eq:Reboth})) when only the inner cylinder is rotating,
and when only the outer cylinder is rotating. 
The results are
presented in Figure~\ref{fig:torque8}; the predictions of the model
show excellent agreement with the experimental data, even near onset,
in the case where only the inner cylinder is rotating. The agreement
is also very good (except for the two widest gap widths) in the case
where only the outer cylinder is rotating, except near onset. However,
Taylor reports that the onset of instability in the case where only
the outer cylinder is rotating undergoes hysteresis (where the
turbulent solution can only be accessed through finite-amplitude
perturbations); since our model assumes that the turbulent solution is
chosen whenever it exists, we represent only the turbulent branch of
the hysteresis loop, whereas it appears that Taylor's data follows the
laminar branch up to a critical Reynolds number that may depend on the
amount of noise present in his apparatus (see Schultz-Grunow 1959 for
an assessment of the critical Reynolds number for the persistence of
laminar flow in a noise-free Couette--Taylor system).

\subsubsection{Onset of instability in Couette--Taylor flows}
\label{sec:onset}

Although our model was initially designed for fully developed
turbulent flows at very high Reynolds number, and in fact in the
context of astrophysical magnetohydrodynamics (Ogilvie 2003), we now
show that the addition of the viscous correction terms (see
Section~\ref{sec:model}) is also able to reproduce (qualitatively, and
to some extent also quantitatively) the onset of instability in the
Couette--Taylor system.

The classic work of Taylor (1923) combined experimental and
theoretical studies of the onset of linear instability in the laminar
Couette--Taylor flow for both corotating and counter-rotating
cylinders. Impressive agreement was found between the appearance of
axisymmetric Taylor vortices in the experiments and the occurrence of
an axisymmetric linear instability in the theoretical calculation.  We
have investigated the linear stability of the laminar flow within
the context of our model.  To do this, we linearize the
Reynolds-stress equation about the laminar solution and seek solutions
of the form $R_{ij}=\tilde R_{ij}(r)\,{\rm e}^{st}$.  The linearized
system of ordinary differential equations admits a set of discrete
modes, with the growth rate $s$ appearing as an eigenvalue.  We solve
this system numerically and identify the stability boundary as the
position in the parameter space where the largest eigenvalue passes
through zero.  The results depend on $C_\nu$, but not on $C_1$ or
$C_2$ as these two parameters appear only in nonlinear terms.

\begin{figure}
   \centerline{\epsfysize8cm\epsfbox{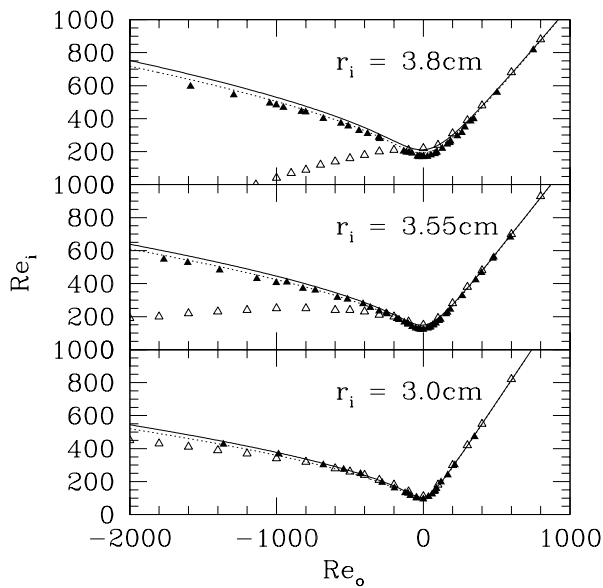}}
   \caption{Comparison between the predictions of our model for linear and
     nonlinear stability of Couette--Taylor flow between two cylinders
     with $r_o = 4.035$ cm and $r_i$ as shown in the plots, and
     Taylor's (1923) experimental data for the onset of instability.
     The black symbols correspond to Taylor's data, the open symbols
     corresponds to the onset of nonlinear instability as calculated
     with our full model for standard parameters.  The solid line
     corresponds to the onset of instability calculated through a
     linear stability analysis of our turbulent model for
     $C_\nu=12.48$ and the dotted lines show the same thing for $C_\nu
     = 11$.}
\label{fig:taylor1923}
\end{figure}

The linear stability boundary predicted by our model is shown in
Figure~\ref{fig:taylor1923} in comparison with Taylor's experimental
results.  For $C_\nu=12.48$ the agreement is quite good (an even
better fit can be obtained for $C_\nu=11$).  We therefore again find
that, although our model is designed principally to describe fully
developed turbulence, it also performs quite well in describing the
onset of instability.  Of course, Taylor vortices themselves are not a
turbulent flow, but our model does not make a clear distinction
between coherent and turbulent flows at relatively low Reynolds
numbers.  The near coincidence between our linear stability results
and Taylor's is not trivial because, unlike him, we do not represent
or solve for the optimal axial wavenumber of the linear disturbance.
  
Also shown in Figure~\ref{fig:taylor1923} is the nonlinear
stability boundary which delimits the region of parameter space in
which our model predicts turbulent solutions to exist. 
The discrepancy with Taylor's results illustrates the fact that a
finite-amplitude instability, apparently not detected in Taylor's
(1923) experiments, may occur in the case of counter-rotating
cylinders. This idea is supported by the results of Coles (1965), who 
reports on the existence of a well-defined hysteresis zone delimited by 
a boundary qualitatively similar to our nonlinear stability boundary.

This boundary turns over and crosses the $Re_i
=0$ axis for finite $Re_o$. The unstable domain thus delimited for
$Re_i<0, Re_o<0$ is the point-symmetric domain to the one identified
as a region of finite-amplitude shear instability in the quadrant
$Re_i>0, Re_o>0$ (see Section~\ref{sec:stabsols} and Figure~\ref{fig:stabbound}).

\subsubsection{Keplerian shear flows}
\label{sec:kep}

An important unsolved problem in astrophysics concerns the
hydrodynamic stability of accretion discs in which gas flows in
circular Keplerian orbits with $\Omega\propto r^{-3/2}$.  Although
magnetohydrodynamic instabilities are known to be effective in
generating turbulent motion and angular momentum transport in discs
that are sufficiently ionized (e.g. Balbus \& Hawley 1998), such
mechanisms probably fail to operate in some important circumstances
such as in very weakly ionized regions of protoplanetary discs.

\begin{figure}
  \centerline{\epsfysize8cm\epsfbox{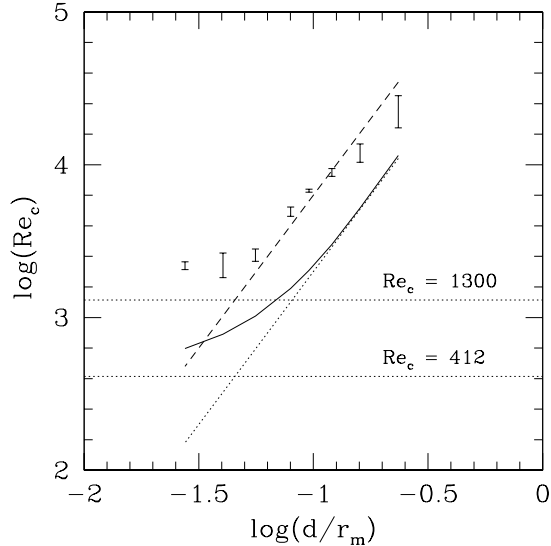}}
  \caption{Critical Reynolds number $Re_c = \min(Re_o)$ 
  for the onset of nonlinear
    instability in the case where the inner cylinder is at rest, as a
    function of gap width ($r_o = 4.05$ cm for all points, and $r_i$
    is varied between $3.2$ cm and $3.94$ cm). The error bars
    reproduce Taylor's own interpretation of his data (1936{\it a})
    with the size of the error bar corresponding to the extent of the
    hysteresis loop. The dashed line is the fit proposed by Richard \&
    Zahn (1999), with $Re_c = 6.3 \times 10^5 (d/r_m)^2$; the solid
    line is our own results, and the inclined dotted line is a fit for
    the wide-gap limit with $Re_c = 2 \times 10^5 (d/r_m)^2$. The two
    horizontal lines are critical Reynolds numbers for plane Couette
    flow: $Re_c = 1300$ is derived from Dauchot \& Daviaud's
    (1995) experiments, and $Re_c = 412$ is the prediction of our
    model.}
\label{fig:recrit}
\end{figure}

Recently, Richard \& Zahn (1999) suggested, not unreasonably, that the
stability of Keplerian flows might be deduced from the results of
wide-gap Couette--Taylor experiments. They extract from Taylor's
(1936{\it a}) and Wendt's (1933) data that the critical Reynolds
number for instability, in the case where the inner cylinder is at
rest, varies roughly as $d^2/r_m^2$ for wide gaps. From this result
they deduce that there must exist a local Reynolds number for rotating
shear flows
\begin{equation}
Re_{RZ} = \frac{r^3}{\nu} \left|\frac{\ptl \Omega}{\ptl r}\right|
\end{equation}
with a critical value $Re_{c,RZ} \simeq 6.3 \times 10^5$ for
instability (see Figure~\ref{fig:recrit}). If such an abstraction of
the Couette--Taylor experiments to Keplerian flows is indeed
justified, this criterion would suggest that Keplerian discs (which
typically have Reynolds numbers many orders of magnitude larger than
this critical value) are indeed likely to be turbulent.
Numerical solutions of our model near the onset of nonlinear
instability in the case where the inner cylinder is at rest also
reveal that the critical Reynolds number for instability varies as
$d^2/r_m^2$ for wide gaps, although the proportionality constant is
lower (see Figure~\ref{fig:recrit}).  The same argument proposed by
Richard \& Zahn (1999) applied to our results would therefore also
suggest that Keplerian shear flows should be unstable.  However, a
local analysis of our model predicts instability for Keplerian shear
flows in the limit of large Reynolds numbers only when $C_2/C_1 > 8/3$
(Ogilvie 2003), which is not the case for the standard model
parameters chosen in this numerical experiment.  Hence, it is not
clear that a generalization between Couette--Taylor results with a
stationary inner cylinder and Keplerian flows can be made.

\begin{figure}
  \centerline{\epsfysize8cm\epsfbox{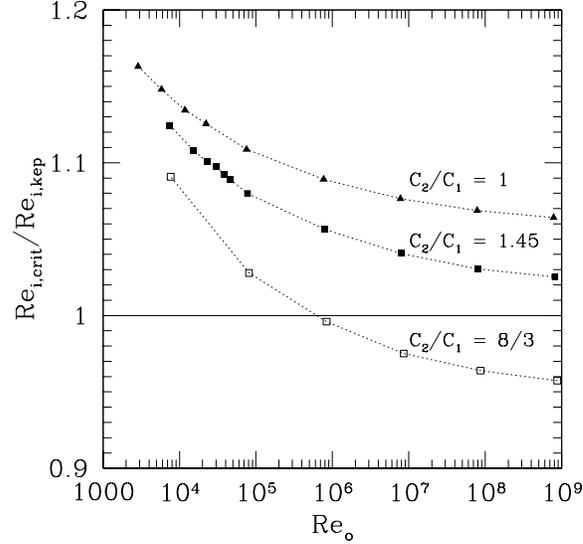}}
  \caption{Predictions of the model for the onset of instability in a
    Couette--Taylor system identical to that of Richard (2001) with
    $r_o = 5$ cm, $r_i = 3.5$ cm. For each outer cylinder rotation
    rate (represented by its Reynolds number $Re_o$), the curves
    represent the ratio of the critical Reynolds number $Re_{i,crit}$
    for instability to the Reynolds number corresponding to a
    Keplerian rotation rate for the inner cylinder $Re_{i,kep} = Re_o
    (r_o/r_i)^{1/2}$.
}
\label{fig:kep}
\end{figure}

More generally, it is not clear that stability results for
wall-bounded flows can be applied to unbounded flows. The instability
may be triggered precisely by the presence of the boundaries (both the
side walls, through the non-local effect of a redistribution of the
shear profile between the cylinders, and the bottom boundary, through
contaminating Ekman flows). For instance, in Figure~\ref{fig:kep} we
explore the stability of Couette--Taylor systems with inner and outer
cylinders in Keplerian ratios ($\Omega_i r_i^{3/2}= \Omega_o
r_o^{3/2}$). As mentioned earlier, if walls were absent and the shear
was everywhere Keplerian, the local analysis of our model would only
predict instability if $C_2/C_1 > 8/3$. However, we find that for
large enough Reynolds numbers, instability can be found for ratios of
$C_2/C_1 < 8/3$ in a wall-bounded experiment.  This behaviour is
possible because the dimensionless Rayleigh discriminant $\phi$ of the
laminar solution is not uniform.

\section{Discussion}

In this paper we have investigated the nonlinear dynamics of turbulent
shear flows, with and without rotation, in the context of a simple but
physically motivated closure of the equation governing the evolution
of the Reynolds stress tensor.  In order to permit a detailed
exploration of the nonlinear behaviour and to emphasize the physical
interpretation of the dynamics, the approach we have taken differs
from that of the conventional closure models used in engineering
applications.  We have not developed a model of great algebraic
sophistication and attempted to fit the large number of parameters
therein by applying a restricted class of constraints.  Instead, we
have adopted a minimal closure of the Reynolds-stress equation in
which the modelled nonlinear terms have a clear interpretation and are
as few in number as is compatible with the physical requirements.

Our model, equation (\ref{drij}), retains the exact form of the linear
terms representing the advection of the turbulent fluctuations by the
mean flow, their interaction with the mean velocity gradient and the
viscous diffusion of the Reynolds stress, while using a minimal set of
algebraic terms with three dimensionless parameters to represent
dissipation through a turbulent cascade (with parameter $C_1$) and
through direct viscous damping (parameter $C_\nu$), as well as the
tendency to return to isotropy (parameter $C_2$).

In a local analysis of homogeneous shear turbulence with or without
rotation (Section~\ref{sec:localhom}), our closure model reduces to an
autonomous nonlinear dynamical system whose fixed points, either
stable or unstable, represent the laminar state and any statistically
steady turbulent states.  We find that the behaviour of the system
depends on the Rayleigh discriminant (defined by equation
(\ref{eq:raydisc})) of the rotating shear flow.  The model predicts
that Rayleigh-unstable flows become turbulent at sufficiently large
Reynolds number through a linear instability associated with a
supercritical (or, rarely, subcritical) bifurcation.  Flows that are
Rayleigh-stable by a sufficiently large margin are predicted not to
support sustained turbulence however large the Reynolds number.  This
behaviour is naturally consistent with Rayleigh's stability criterion.

Non-rotating (Rayleigh-neutral) shear flows and those that are
Rayleigh-stable by a sufficiently small margin can become turbulent
through a nonlinear instability associated with a subcritical
bifurcation from infinite Reynolds number.  In the non-rotating case
the laminar state admits algebraically growing infinitesimal
disturbances that are damped only on the viscous timescale $Re/S$.
The nonlinear terms of the model allow perturbations of finite
amplitude to be sustained and the system evolves to a non-trivial
state of statistically steady turbulence.  This behaviour of the model
strongly resembles the theory of subcritical transition to turbulence,
developed by Trefethen et al. (1993) and others, involving the
transient amplification of disturbances by a non-normal operator and a
cooperative nonlinear feedback.  The closure model that we work with
has the advantage of being able to represent the final turbulent
outcome of the transition process.

The analysis of Reynolds-stress models in homogeneous shear flow in
terms of a nonlinear dynamical system is not unique to our work (see,
e.g., Speziale, Gatski \& Mac Giolla Mhuiris 1990).  However, the
simplicity of our model permits an exhaustive study of its dynamical
properties and, by including the effects of a finite Reynolds number,
we are able to make a connection with the theory of subcritical
transition to turbulence in which the laminar state has a basin of
attraction that diminishes as the Reynolds number is increased.
Similar techniques of analysis could of course be applied to more
sophisticated closure models and we believe that our findings are to
some extent generic.

The turbulent solutions are anisotropic as a result of shear and
rotation, and in the limit of large Reynolds number the shear stress
behaves as in Prandtl's mixing-length theory, but with a prefactor
that depends on the Rayleigh discriminant (see equation
(\ref{eq:rxy_local})). As such, our model naturally captures the
reduction, and eventually suppression, of the turbulent energy
dissipation for rapidly rotating flows (Speziale et al. 1998).

When applied to wall-bounded turbulent shear flows
(Section~\ref{sec:univ_boundlayer}), the model predicts the occurrence
of a universal velocity profile close to a wall at large Reynolds
number.  Outside the viscous sublayer, this profile has the
logarithmic form predicted by Prandtl's mixing-length theory, and we
derive two accurate constraints on the three parameters from matching
the most recent experimental data (Zagarola \& Smits 1998). 

We have also investigated in some detail the predictions of the model
for the occurrence and the characteristics of turbulent states in
Couette--Taylor flow without end-effects (Section~\ref{sec:ctf}).
Here, depending on the ratios of the radii and angular velocities of
the two cylinders, the distribution of the Rayleigh discriminant of
the laminar solution may be such that a local analysis would predict
either linear instability or nonlinear instability or complete stability
in different regions of the flow.  Furthermore, once turbulence sets
in, the angular velocity distribution and the corresponding Rayleigh
discriminant are significantly modified from those of the laminar
solution.  Therefore a wide variety of behaviour is possible,
including the existence of mixed states, in which the turbulence is
localized.  It is worth noting that although we have restricted the
present analysis to solutions of maximal symmetry in the
Couette--Taylor system, our model may admit classes of more general
solutions. For instance, relaxing the assumption of azimuthal and
axial translational symmetry could in principle help explain observed
phenomena such as spiral turbulence (Coles 1965; Hegseth et al. 1989)
in which regions of laminar and turbulent flows coexist separated by a
helical interface.

By fitting the remaining parameter of our model, we are able to
account quite well for the qualitative behaviour and quantitative
torque measurements in historical experiments on Couette--Taylor flow
by Wendt (1933) and Taylor (1936{\it a}), which have not been
superseded.  As an unexpected bonus, the model captures reasonably
accurately the appearance of Taylor vortices at the onset of linear
instability.

It is appropriate to discuss at this point some of the implications of
the parameterization used in our model. The ratio $C_2/C_1$ represents
the propensity of the turbulence to return to isotropy.  In a local
analysis, it is found to determine the critical value of the
dimensionless Rayleigh discriminant for which high-Reynolds-number
turbulence can be sustained.  Why these properties should be related
can be explained with reference to the system of equations
(\ref{dynamical_system}).  When the Rayleigh discriminant
$\Phi=2\Omega(2\Omega-S)$ is positive, the terms $4\Omega R_{xy}$ and
$2(S-2\Omega)R_{xy}$ in the equations for $R_{xx}$ and $R_{yy}$ have
opposite signs.  Therefore either $R_{xx}$ or $R_{yy}$ lacks a
positive source, and the turbulence must decay, unless the $C_2$ term
comes into play.  For a given positive Rayleigh discriminant, the
isotropizing tendency must be sufficiently great if the turbulence is
to be sustained.

In this work, although we have proposed values of the coefficients
$C_1, C_2$ and $C_\nu$ after fitting experimental data, these values
are only tentative and approximate and we do not claim that such a
simple model can provide great quantitative accuracy in comparison
with currently available closure models (cf. Choi \& Lumley 2001) .
In particular, the ratio $C_2/C_1$ is only weakly constrained through
a comparison with Wendt's (1933) data, with a wide plausible range of
roughly 0.4 to 1.

We draw attention again to the important problem of the hydrodynamic
stability of circular Keplerian motion in astrophysical accretion
discs, in which the angular velocity profile $\Omega\propto r^{-3/2}$
is enforced by gravitational dynamics, not through the boundaries.
While Richard \& Zahn (1999) sought to apply the findings of
Couette--Taylor experiments to accretion discs, our investigation of
turbulent Couette--Taylor flows suggests that caution is required in
making such associations.  According to our model (taking $C_2/C_1 <
8/3$), Keplerian rotation is likely not to support statistically
steady turbulence in a local analysis, and may be nonlinearly stable
no matter how large the Reynolds number. We also find that this does
not contradict the experimental finding that Couette--Taylor flow with
a stationary inner cylinder becomes turbulent at large Reynolds
number, and is even consistent with the possibility that a wide-gap
Couette--Taylor flow with the cylinders in a Keplerian ratio may be
turbulent.  In addition, Couette--Taylor experiments are always
contaminated to some degree by end-effects.  We suggest that
Couette--Taylor experiments may be of limited applicability to the
study of the nonlinear stability of Keplerian rotation, and that it
can be instead most usefully addressed in local numerical models such
as that of the shearing box (e.g. Balbus \& Hawley 1998), which are
free from end-effects and also from the type of radial boundary
conditions that induce boundary layers.  To date, no instability has
been found in such models, and it would be valuable to test this
finding to very high Reynolds numbers.

Finally, we emphasize that the philosophy behind the construction of
simple turbulence models such as the one adopted here is applicable to
a range of more complex problems such as magnetohydrodynamic
turbulence (Ogilvie 2003), convection, or mixing in stratified shear
flows. Such extension is the subject of current investigations.

\begin{acknowledgements}
  PG acknowledges the support of New Hall and PPARC.  GIO acknowledges
  the support of the Royal Society through a University Research
  Fellowship.
\end{acknowledgements}

\appendix
\section{Model equations in a general orthogonal curvilinear
coordinate system}

Using Batchelor's (1967) notation, the equivalent of equation
(\ref{drij_old}) for the evolution of the Reynolds stress tensor in
general orthogonal curvilinear coordinates is
\begin{eqnarray}
\frac{\ptl  R_{ij}}{\ptl t} &+& \sum_k \left[\frac{\ou_k}{h_k} \frac{\ptl R_{ij}}{\ptl x_k} + \frac{R_{kj}}{h_k} \frac{\ptl  \ou_i}{\ptl x_k} +  \frac{R_{ik}}{h_k} \frac{\ptl  \ou_j}{\ptl x_k} + \frac{\ou_i R_{kj}}{h_i h_k} \frac{\ptl h_i}{\ptl x_k} + \frac{\ou_k R_{ij}}{h_i h_k} \frac{\ptl h_i}{\ptl x_k}  \right. \nonumber \\ &&\quad \left. +\frac{\ou_j R_{ik}}{h_j h_k} \frac{\ptl h_j}{\ptl x_k} +  \frac{\ou_k R_{ij}}{h_j h_k} \frac{\ptl h_j}{\ptl x_k}-  2  \frac{\ou_k R_{kj}}{h_i h_k} \frac{\ptl h_k}{\ptl x_i} - 2 \frac{\ou_k R_{ki}}{h_j h_k} \frac{\ptl h_k}{\ptl x_j} \right] \nonumber \\ &=& -C_1 L^{-1} R^{1/2} R_{ij} -C_2 L^{-1} R^{1/2} (R_{ij}-\textstyle{\frac{1}{3}} R \delta_{ij}) \mbox{   ,}
\end{eqnarray}
where $(h_1,h_2,h_3)$ is $(1,1,1)$ for Cartesian coordinates
$(x,y,z)$, or $(1,r,1)$ for cylindrical coordinates $(r,\phi,z)$.

In the Cartesian case, the viscous correction terms follow from the
decomposition
\begin{equation}
\nu (u'_{i,kk} u'_j + u'_{j,kk} u'_i) = \nu ((u'_iu'_j)_{,kk} - 2 
u'_{i,k}u'_{j,k})\mbox{   ,}
\end{equation}
where the first term on the right-hand side describes a viscous
diffusion of the stresses and the second term describes a direct decay
of the stresses, which we then model as $- C_{\nu}\nu R_{ij} /L^2$.

In order to obtain the form for the viscous corrections in general 
curvilinear coordinates, we
follow the same method used in the Cartesian case: we isolate from the
original terms $\nu (\grad^2 u')_i u'_j + \nu (\grad^2 u')_j u'_i$ the
tensor decay term $-2\nu(\grad \bu' \grad^T \bu')_{ij}$ (which is the
covariant equivalent of $-2\nu u'_{i,k} u'_{j,k}$), where $\grad \bu'$
is the matrix defined by its columns
\begin{equation}
(\grad \bu') = \left(\frac{1}{h_1} \frac{\ptl \bu'}{\ptl
x_1},\frac{1}{h_2} \frac{\ptl \bu'}{\ptl x_2},\frac{1}{h_3} \frac{\ptl
\bu'}{\ptl x_3}\right)\mbox{   .}
\end{equation}
In this expression, $\bu' = u'_1 \be_1 + u'_2 \be_2 + u'_3 \be_3$ and
the derivatives of the unit vectors $(\be_1,\be_2,\be_3)$ are given by
Batchelor (1967, p. 598). We model the decay terms as $- C_{\nu}\nu R_{ij}
/L^2$, and keep the remaining diffusion terms unchanged thereby
defining the Laplacian of the second rank tensor $(\grad^2 R)_{ij}$.
Hence with this method the viscous terms in equation (\ref{drij})
are then simply
\begin{equation}
- \frac{C_{\nu}\nu}{L^2} R_{ij} + \nu (\grad^2 R)_{ij} \mbox{   ,}
\end{equation}
with
\begin{eqnarray}
&& (\grad^2 R)_{rr} = \grad^2 R_{rr} - \frac{4}{r^2} \frac{\ptl R_{r\phi}}{\ptl \phi} + \frac{2}{r^2} (R_{\phi\phi} - R_{rr}) \nonumber\mbox{   ,}\\
&& (\grad^2 R)_{\phi\phi}  = \grad^2 R_{\phi\phi} + \frac{4}{r^2} \frac{\ptl R_{r\phi}}{\ptl \phi} + \frac{2}{r^2} (R_{rr} - R_{\phi\phi}) \nonumber\mbox{   ,} \\
&& (\grad^2 R)_{zz} = \grad^2 R_{zz}\nonumber\mbox{   ,} \\
&& (\grad^2 R)_{r\phi} = \grad^2 R_{r\phi} + \frac{2}{r^2} \frac{\ptl}{\ptl \phi }(R_{rr} - R_{\phi\phi}) - \frac{4}{r^2} R_{r\phi} \nonumber \mbox{   ,}\\
&& (\grad^2 R)_{rz} = \grad^2 R_{rz} - \frac{2}{r^2} \frac{\ptl R_{\phi z}}{\ptl \phi} - \frac{R_{rz}}{r^2} \mbox{   ,}\nonumber\\
&& (\grad^2 R)_{\phi z} = \grad^2 R_{\phi z} + \frac{2}{r^2} \frac{\ptl R_{rz}}{\ptl \phi} - \frac{R_{\phi z}}{r^2} \mbox{   ,}
\end{eqnarray} 
in cylindrical coordinates for instance. 

For an axisymmetric flow with translational symmetry in $z$ and
$\bar{\bu} = r\Omega(r)\,\be_\phi$ in a cylindrical geometry the
evolution equations for the Reynolds stress tensor are
\begin{eqnarray}
\frac{\ptl R_{rr}}{\ptl t} - 4\Omega R_{r\phi} &=& -\frac{C_1+C_2}{L} R^{1/2} R_{rr} + \frac{C_2}{3L} R^{3/2} - \frac{\nu C_{\nu}}{L} R_{rr} + \nu (\grad^2 R)_{rr}\mbox{   ,} \nonumber \\
\frac{\ptl R_{\phi\phi}}{\ptl t} + 2(2\Omega-S) R_{r \phi} &=&  -\frac{C_1+C_2}{L} R^{1/2} R_{\phi\phi} + \frac{C_2}{3L} R^{3/2} - \frac{\nu C_{\nu}}{L} R_{\phi\phi} + \nu (\grad^2 R)_{\phi\phi} \mbox{   ,} \nonumber \\
\frac{\ptl R_{zz}}{\ptl t} &=& - \frac{C_1+C_2}{L} R^{1/2} R_{zz} + \frac{C_2}{3L} R^{3/2}  - \frac{\nu C_{\nu}}{L} R_{zz} + \nu (\grad^2 R)_{zz} \mbox{   ,} \nonumber \\
\frac{\ptl R_{r\phi}}{\ptl t} + (2\Omega-S) R_{rr} - 2 \Omega R_{\phi\phi} &=& -\frac{C_1+C_2}{L} R^{1/2}R_{r\phi} - \frac{\nu C_{\nu}}{L} R_{r\phi} + \nu (\grad^2 R)_{r\phi} \mbox{   ,}\nonumber \\
\frac{\ptl R_{rz}}{\ptl t} - 2 \Omega R_{\phi z} &=& -\frac{C_1+C_2}{L} R^{1/2}R_{rz} - \frac{\nu C_{\nu}}{L} R_{rz} + \nu (\grad^2 R)_{rz} \mbox{   ,}\nonumber \\
\frac{\ptl R_{\phi z}}{\ptl t} + (2\Omega-S) R_{rz} &=& -\frac{C_1+C_2}{L} R^{1/2}R_{\phi z} - \frac{\nu C_{\nu}}{L} R_{\phi z} + \nu (\grad^2 R)_{\phi z}\mbox{   ,}
\label{eq:Reqcyl}
\end{eqnarray}
where 
\begin{equation}
 S = -r\frac{\dd \Omega}{\dd r}\mbox{   .}
\label{eq:Sdef}
\end{equation} 

\section{Asymptotic solution of the turbulent Couette--Taylor flow for 
  large Reynolds number and large Rossby number}

In the limits of large Rossby number (as expected for a small gap) and
large Reynolds number, the angular momentum equation (\ref{eq:am}) can
be approximated (to first order in $Ro^{-1}$) by
\begin{equation}
r^2 \kappa^2  L^2 S|S| + 2 r^2 \kappa^2 L^2 \Omega |S| \frac{9 C_1}{C_2} = \frac{T}{2\pi h \rho}\mbox{   ,}
\end{equation}
where $\kappa$ is the von K\'arm\'an constant defined in equation
(\ref{eq:vk}).  This provides a quadratic equation for $S$ which can
be inverted, and yields (in the same limit)
\begin{equation}
S = -r \frac{\dd \Omega}{\dd r} = - \frac{9 C_1}{ C_2} \Omega + {\rm sign}(S) \frac{1}{rL(r)\kappa} \sqrt{\frac{|T|}{2\pi h \rho}}\mbox{   .}
\end{equation}
This equation must be integrated separately in each intervals
$(r_i,r_m]$ and $[r_m,r_o)$: in $(r_i,r_m]$,
\begin{equation}
\Omega(r_m) r_m^{-9C_1/C_2} - \Omega(r)r^{-9C_1/C_2} = -\frac{{\rm sign}(S)}{\kappa} \sqrt{\frac{|T|}{2 \pi h \rho}}  \int_r^{r_m} \frac{r'^{-2-9C_1/C_2}}{r'-r_i} \dd r' \mbox{   ,}
\label{eq:omegain}
\end{equation}
and in $[r_m,r_o)$,
\begin{equation}
\Omega(r) r^{-9C_1/C_2} - \Omega(r_m)r_m^{-9C_1/C_2} = -\frac{{\rm sign}(S)}{\kappa} \sqrt{\frac{|T|}{ 2\pi h \rho}}  \int_{r_m}^{r} \frac{r'^{-2-9C_1/C_2}}{r_o-r'} \dd r' \mbox{   .}
\label{eq:omegaout}
\end{equation}
Let $\alpha = 9C_1/C_2 + 2$. The integrals in equations
(\ref{eq:omegain}) and (\ref{eq:omegaout}) have a logarithmic
singularity as $r \rightarrow r_i$ and $r \rightarrow r_o$, which can
be isolated as
\begin{equation}
\int_r^{r_m} \frac{r'^{-\alpha}}{r'-r_i} \dd r' = f(r;r_m,r_i,\alpha) - r_i^{-\alpha} \ln \left(\frac{ r-r_i}{r_m-r_i}\right)\mbox{   .}
\end{equation}
This defines the function $f$
uniquely. The logarithmic singularity naturally matches onto the
boundary layer solutions near the walls.

We use the results of Section~\ref{sec:univ_boundlayer} to write the
boundary layer solution explicitly. Near $r=r_i$, but outside the
viscous sublayer,
\begin{equation}
\Omega(r) = \Omega_i - \frac{{\rm sign}(S)}{r_i^2}  \sqrt{ \frac{|T|}{2\pi h \rho}} \left[ v_0(C_1,C_2,C_{\nu}) + \frac{1}{\kappa} \ln \left( \frac{r-r_i}{r_i \nu} \sqrt{ \frac{|T|}{2\pi h \rho}} \right) \right], 
\label{eq:blin}
\end{equation} 
whereas near $r=r_o$, 
\begin{equation}
\Omega(r) = \Omega_o + \frac{{\rm sign}(S)}{r_o^2}  \sqrt{ \frac{|T|}{2\pi h \rho}} \left[ v_0(C_1,C_2,C_{\nu}) + \frac{1}{\kappa} \ln \left( \frac{r_o-r}{r_o \nu} \sqrt{ \frac{|T|}{2\pi h \rho}} \right) \right] \mbox{   .}
\label{eq:blout}
\end{equation} 
Matching the inner (\ref{eq:blin}), (\ref{eq:blout}) and outer
(\ref{eq:omegain}), (\ref{eq:omegaout}) solutions near the walls, and
continuity across the mid-point $r_m$ provides an equation for the
torque, which extends the Prandtl--von K\'arm\'an skin-friction law
for Couette--Taylor flows:
\begin{equation}
\frac{1}{\sqrt{C_f}} = N \log_{10} (Re\sqrt{C_f}) + M\mbox{   ,}
\label{eq:skinfriction}
\end{equation}
where we have defined\footnote{Other authors who use a different
  definition of the Reynolds number $Re^*$ obtain skin-friction law
  coefficients $M^*$ and $N^*$ which are related to ours through the
  expression $M^* = (Re^*/Re) M$ and $N^* = (Re^*/Re) N$.}  $Re =
|\Omega_i- \Omega_o|(r_o^2-r_i^2)/(2\nu)$, $C_f = T/Re^2 \rho h \nu^2
$ and the friction law coefficients $M$ and $N$ as
\begin{eqnarray}
&&N = \frac{{\rm sign}(S)}{2\kappa \sqrt{2\pi}} \frac{|\Omega_i - \Omega_o |(r_o^2-r_i^2)}{\Omega_i r^{2-\alpha}_i - \Omega_o r_o^{2-\alpha}} \left(r_i^{-\alpha} + r_o^{-\alpha}\right) \ln 10 \mbox{   ,}\nonumber \\
&& M = \frac{{\rm sign}(S)}{2\kappa\sqrt{2\pi }} \frac{|\Omega_i- \Omega_o |( r_o^2-r_i^2)}{\Omega_i r^{2-\alpha}_i - \Omega_o r_o^{2-\alpha}} \left[ \left(r_i^{-\alpha} + r_o^{-\alpha}\right) \left( \ln \frac{r_o-r_i}{\sqrt{8\pi}}  + \kappa v_0 (C_1,C_2,C_{\nu})\right) \right. \nonumber \\ &&\qquad\quad \left. + \left(f(r_i;r_m,r_i,\alpha)+f(r_o;r_m,r_o,\alpha)\right) - \left(\frac{ \ln r_i}{r_i^{\alpha}} + \frac{ \ln r_o}{r_o^{\alpha}} \right) \right]\mbox{   .}
\label{eq:M&N}
\end{eqnarray}
Note that in the limit where the contribution from rotation on the
Reynolds stresses is neglected ($Ro^{-1} = 0$), the solution in the
bulk of the fluid can be written out as equations (\ref{eq:omegain})
and (\ref{eq:omegaout}) with $\alpha = 2$ and
\begin{equation}
f(r;r_m,r_i,2) = \frac{1}{r_i^2} \left( \frac{1}{r_m} - \frac{1}{r} +  \ln r - \ln r_m \right)\mbox{   .}
\end{equation}

\begin{figure}
   \centerline{\epsfysize8cm\epsfbox{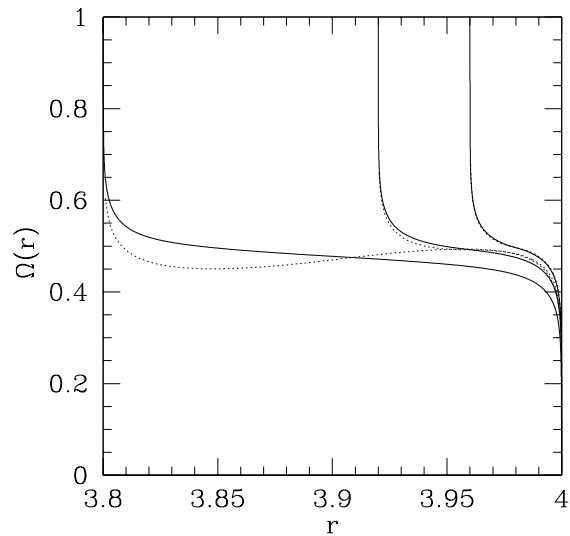}}
   \caption{Comparison between complete numerical solution (solid
     line) and asymptotic analytical solution (dotted line) for three
     gap widths $d/r_o = 0.05$, 0.02 and 0.01 respectively, for a
     Reynolds number of the flow $Re = 10^6$.}
\label{fig:compareVcyl}
\end{figure}

In Figure~\ref{fig:compareVcyl} we compare the velocity profiles
obtained by numerical integration to those derived from asymptotic
analysis, for three different gap widths. We find that the asymptotics
only provide accurate results for $d/r_o \le 0.02$. This somewhat
disappointing range of applicability of the $Ro\gg1$ asymptotic
analysis is due to the great efficiency with which turbulence
redistributes the shear, which reduces the Rossby number in
the interior of the flow compared to that of the laminar solution.


\begin{thebibliography}{}
  
\bibitem[Andereck, Liu \& Swinney (1986)]{ALS86}
  \textsc{Andereck, C. D., Liu, S. S. \& Swinney, H. L.} 1986
  Flow regimes in a circular Couette system with independently rotating
  cylinders.
  \textit{J. Fluid Mech.} \textbf{164}, 155--183.

\bibitem[Baggett \& Trefethen (1997)]{BT97}
  \textsc{Baggett, J. S. \& Trefethen, L. N.} 1997
  Low-dimensional models of subcritical transition to turbulence.
  \textit{Phys. Fluids} \textbf{9}, 1043--1053.

\bibitem[Balbus \& Hawley (1998)]{BH98}
  \textsc{Balbus, S. A. \& Hawley, J. F.} 1998
  Instability, turbulence, and enhanced transport in accretion disks.
  \textit {Rev. Mod. Phys.} \textbf{70}, 1--53.

\bibitem[Batchelor (1967)]{B67}
  \textsc{Batchelor, G. K.} 1967
  \textit{An Introduction to Fluid Dynamics.}  Cambridge University Press.


\bibitem[Boussinesq (1877)]{B87}
  \textsc{Boussinesq, J.} 1877
  Essai sur la th\'eorie des eaux courantes.
  \textit{M\'em. Acad. Sci. Paris} \textbf{23}, 1--660.

\bibitem[Butler \& Farrell (1992)]{BF92}
  \textsc{Butler, K. M. \& Farrell, B. F.} 1992
  Three-dimensional optimal perturbations in viscous shear flow.
  \textit{Phys. Fluids} A \textbf{4}, 1637--1650.

\bibitem[Choi \& Lumley (2001)]{CL01}
  \textsc{Choi, K. S. \& Lumley, J. L.} 2001
  The return to isotropy of homogeneous turbulence.
  \textit{J. Fluid Mech.} \textbf{436}, 59--84.

\bibitem[Coles (1965)]{C65}
  \textsc{Coles, D.} 1965  
  Transition in circular Couette flow.  
  \textit{J. Fluid Mech.} \textbf{21}, 385--425.

\bibitem[Dauchot \& Daviaud (1995)]{DD95}
  \textsc{Dauchot, O. \& Daviaud, F.} 1995
  Finite amplitude perturbation and spot growth mechanism 
  in plane Couette flow.
  \textit{Phys. Fluids} A \textbf{7}, 335--343.

\bibitem[Drazin \& Reid (1981)]{DR81}
  \textsc{Drazin, P. G. \& Reid, W. H.} 1981
  \textit{Hydrodynamic Stability.}  Cambridge University Press.

\bibitem[Frank, King \& Raine (2002)]{FKR02}
  \textsc{Frank, J., King, A. \& Raine, D.} 2002
  \textit{Accretion Power in Astrophysics,} 3rd edn. Cambridge
  University Press.

\bibitem[Grossmann (2000)]{G00}
  \textsc{Grossmann, S.} 2000
  The onset of shear flow turbulence.
  \textit{Rev. Mod. Phys.} \textbf{72}, 603--618.

\bibitem[Hegseth, Andereck, Hayot, \& Pomeau (1989)]{HAHP89}
   \textsc{Hegseth, J. J., Andereck, C. D., Hayot, F. \& Pomeau, Y.} 1989
   Spiral turbulence and phase dynamics.
   \textit{Phys. Rev.} E \textbf{62}, 257--260

\bibitem[Lathrop, Fineberg \& Swinney (1992)]{LFS92}
  \textsc{Lathrop, D. P., Fineberg, J. \& Swinney, H. L.} 1992
  Transition to shear-driven turbulence in Couette--Taylor flow.
  \textit{Phys. Rev.} A \textbf{46}, 6390--6405.

\bibitem[Launder, Reece \& Rodi (1975)]{LRR75}
  \textsc{Launder, B. E., Reece, G. J. \& Rodi, W.} 1975
  Progress in the development of a Reynolds stress turbulence closure.
  \textit{J. Fluid Mech.} \textbf{68}, 537--566.

\bibitem[Lewis \& Swinney (1999)]{LS99}
  \textsc{Lewis, G. S. \& Swinney, H. L.} 1999
  Velocity structure functions, scalings, and transition in 
  high-Reynolds-number Couette--Taylor flow.
  \textit{Phys. Rev.} E \textbf{59}, 5457--5467.

\bibitem[McKeon et al. (2003)]{MK03}
  \textsc{McKeon, B. J., Jiang, W., Morrison, J. F. \& Smits, A. J.} 2003
  Pitot probe corrections in fully developed turbulent pipe flow. 
  \textit{Meas. Sci. Technol.} \textbf{14}, 1449--1458 

\bibitem[Nagata (1990)]{N90}
  \textsc{Nagata, M.} 1990
  Three-dimensional finite-amplitude solutions in plane Couette flow:
  bifurcation from infinity.
  \textit{J. Fluid Mech.} \textbf{217}, 519--527.

\bibitem[Ogilvie (2003)]{O03}
  \textsc{Ogilvie, G. I.} 2003
  On the dynamics of magnetorotational turbulent stresses.
  \textit{Mon. Not. R. Astron. Soc.} \textbf{340}, 969--982.

\bibitem[Prandtl (1925)]{P25}
  \textsc{Prandtl, L.} 1925
  Bericht \"uber Untersuchungen zur ausgebildeten Turbulenz.
  \textit{Z. Angew. Math. Mech.} \textbf{5}, 136--139.

\bibitem[Pringle (1981)]{P81}
  \textsc{Pringle, J. E.} 1981
  Accretion discs in astrophysics.
  \textit{Annu. Rev. Astron. Astrophys.} \textbf{19}, 137--162.

\bibitem[Pumir (1996)]{P96}
  \textsc{Pumir, A.} 1996
  Turbulence in homogeneous shear flows.
  \textit{Phys. Fluids} \textbf{8}, 3112--3127.

\bibitem[Rayleigh (1917)]{R17}
  \textsc{Rayleigh, Lord} 1917
  On the dynamics of revolving fluids.
  \textit{Proc. R. Soc. Lond.} A \textbf{93}, 148--154.

\bibitem[Reichardt (1940)]{R40}
  \textsc{Reichardt, H.} 1940
  Die W\"arme\"ubertragung in turbulenten Reibungsschichten. 
  \textit{Z. Angew. Math. Mech.} \textbf{20}, 297--328.

\bibitem[Reynolds (1895)]{R95}
  \textsc{Reynolds, O.} 1895
  On the dynamical theory of incompressible viscous fluids and the
  determination of the criterion.
  \textit{Phil. Trans. R. Soc. Lond. A} \textbf{186}, 123--164.

\bibitem[Richard (2001)]{R01}
  \textsc{Richard, D.} 2001
  Instabilit\'es hydrodynamiques dans les \'ecoulements en rotation
  diff\'erentielle.
  PhD dissertation, Universit\'e Paris 7 Denis Diderot.

\bibitem[Richard \& Zahn (1999)]{RZ99}
  \textsc{Richard, D. \& Zahn, J.-P.} 1999
  Turbulence in differentially rotating flows: what can be learned
  from the Couette--Taylor experiment.
  \textit{Astron. Astrophys.} \textbf{347}, 734--738.

\bibitem[Richard et al. (2001)]{RDDZ01}
  \textsc{Richard, D., Dauchot, O., Daviaud, F. \& Zahn, J.-P.} 2001
  Subcritical instabilities of astrophysical interest in
  Couette--Taylor systems.
  In \textit{Proc. 12th Couette--Taylor Workshop,
  Evanston, USA}.

\bibitem[Rogallo (1981)]{R81}
  \textsc{Rogallo, R. S.} 1981
  Numerical experiments in homogeneous turbulence.
  \textit{NASA Tech. Mem.} TM-81315.

\bibitem[Rotta (1951)]{R51}
  \textsc{Rotta, J. C.} 1951
  Statistische Theorie nichthomogener Turbulenz.
  \textit{Z. Phys.} \textbf{129}, 547--572.

\bibitem[Schlichting (1979)]{S79}
  \textsc{Schlichting, H.} 1979
  \textit{Boundary-layer Theory}, 7th ed.  McGraw-Hill.

\bibitem[Schultz-Grunow (1959)]{S59}
  \textsc{Schultz-Grunow, F.} 1959
  Zur Stabilit\"at der Couette-Str\"omung.
  \textit{Z. Angew. Math. Mech.} \textbf{39}, 101--110.
  
\bibitem[Sj\"ogren \& Johansson (2000)]{SJ00}
  \textsc{Sj\"ogren, T. \& Johansson A. V.} 2000
  Development and calibration of algebraic nonlinear models for terms
  in the Reynolds stress transport equations.
  \textit{Phys. Fluids} \textbf{12}, 1554--1572.

\bibitem[Speziale (1991)]{S91}
  \textsc{Speziale, C. G.} 1991
  Analytical methods for the development of Reynolds-stress closures
  in turbulence.
  \textit{Annu. Rev. Fluid Mech.} \textbf{23}, 107--157.

\bibitem[Speziale, Gatski \& Mac Giolla Mhuiris (1990)]{SGM90}
  \textsc{Speziale, C. G., Gatski, T. B. \& Mac Giolla Mhuiris, N.} 1990
  A critical comparison of turbulence models for homogeneous shear flows
  in a rotating frame.
  \textit{Phys. Fluids A} \textbf{2}, 1678--1684.

\bibitem[Speziale et al. (1998)]{S98}
  \textsc{Speziale, C. G., Younis, B. A., Rubinstein, R. \& Zhou, Y.} 1998
   On consistency conditions for rotating turbulent flows. 
  \textit{Phys. Fluids} \textbf{10}, 2108--2110.

\bibitem[Tassoul (1978)]{T78}
  \textsc{Tassoul, J.-L.} 1978
  \textit{Theory of Rotating Stars.}  Princeton University Press.

\bibitem[Taylor (1923)]{T23}
  \textsc{Taylor, G. I.} 1923
  Stability of a viscous liquid contained between two rotating cylinders.
  \textit{Phil. Trans. R. Soc. Lond. A} \textbf{223}, 289--343.

\bibitem[Taylor (1936{\it a})]{T36a}
  \textsc{Taylor, G. I.} 1936{\it a}
  Fluid friction between rotating cylinders.  I. Torque measurements.
  \textit{Proc. R. Soc. Lond. A}, \textbf{157}, 546--564.

\bibitem[Taylor (1936{\it b})]{T36b}
  \textsc{Taylor, G. I.} 1936{\it b}
  Fluid friction between rotating cylinders. II. Distribution of
  velocity between concentric cylinders when outer one is rotating and
  inner one is at rest.
  \textit{Proc. R. Soc. Lond. A}, \textbf{157}, 565--578.

\bibitem[Thomson (1887)]{T87}
  \textsc{Thomson, W. (Lord Kelvin)} 1887
  Stability of fluid motion --- rectilineal motion of viscous fluid
  between two parallel planes.
  \textit{Phil. Mag.} \textbf{24} (5), 188--196.

\bibitem[Trefethen et al. (1993)]{TTRD93}
  \textsc{Trefethen, L. N., Trefethen, A. E., Reddy, S. C. \&
  Driscoll, T. A.} 1993
  Hydrodynamic stability without eigenvalues.
  \textit{Science} \textbf{261}, 578--584.

\bibitem[Waleffe (1997)]{W97}
  \textsc{Waleffe, F.} 1997
  On a self-sustaining process in shear flows.
  \textit{Phys. Fluids} \textbf{9}, 883--900.

\bibitem[Wendt (1933)]{W33}
  \textsc{Wendt, F.} 1933
  Turbulente Str\"omungen zwischen zwei rotierenden konaxialen Zylindern.
  \textit{Ing. Archiv} \textbf{4}, 577--595.

\bibitem[Zagarola \& Smits (1998)]{ZS98}
  \textsc{Zagarola, M. V. \& Smits A. J.} 1998
  Mean-flow scaling of turbulent pipe flow.
  \textit{J. Fluid Mech.} \textbf{373}, 33--79.

\end{thebibliography}
\end{document}